\documentclass[aps,prb,reprint,superscriptaddress]{revtex4-2}

\usepackage[utf8]{inputenc}
\usepackage{subfigure}
\usepackage{color}
\usepackage{xcolor}
\usepackage{graphicx}
\usepackage{dcolumn}
\usepackage{bm}
\usepackage{caption}
\usepackage{bbm}

\usepackage{amsmath,leftidx}
\usepackage{times}
\usepackage{CJK}
\usepackage{color,colortbl}
\usepackage[normalem]{ulem}

\usepackage{graphicx}

\newcommand{\bfk}{{\bf k}}

\newcommand{\bfq}{{\bf q}}
\newcommand{\bfr}{{\bf r}}

\newcommand{\be}{\begin{eqnarray}}
\newcommand{\ee}{\end{eqnarray}}

\newcommand{\wbe}{\begin{widetext}}
\newcommand{\wee}{\end{widetext}}
\definecolor{amethyst}{rgb}{0.6, 0.4, 0.8}
\definecolor{antiquefuchsia}{rgb}{0.57, 0.36, 0.51}

\begin{document}

\preprint{APS/123-QED}

\title{2D Gapless Topological Superfluids Generated by Pairing Phases}

\author{Jiapei Zhuang}
\affiliation{ Department of Physics, National Tsing Hua University, Hsinchu 30013, Taiwan}

\author{Ching-Yu Huang}
\affiliation{ Department of Applied Physics, Tunghai University, Taichung 40704,Taiwan}

\author{Po-Yao Chang}
\affiliation{ Department of Physics, National Tsing Hua University, Hsinchu 30013, Taiwan}

\author{Daw-Wei Wang}
\affiliation{ Department of Physics, National Tsing Hua University, Hsinchu 30013, Taiwan}
\affiliation{Frontier Center for Theory and Computation, National Tsing Hua University, Hsinchu 30013, Taiwan}
\affiliation{Physics Division, National Center for Theoretical Sciences, Taipei 10617, Taiwan}
\affiliation{Center for Quantum Technology, National Tsing Hua University, Hsinchu 30013, Taiwan}

\date{\today}
\begin{abstract}
We systematically investigate the ground state phase diagram and the finite temperature phase transitions for a Rydberg-dressed Fermi gas loaded in a bilayer optical lattice. When an effective finite-ranged attraction is induced, our self-consistent mean-field calculation shows that the gapped topological ($p$-wave) superfluids in each layer are coupled together by the $s$-wave pairing in an intermediate inter-layer distance with a spontaneously modulated phases between these two order parameters. The obtained ground state is a gapless topological superfluid with quantized topological charges characterizing the gapless points, leading to a zero energy flat band at the edges. Finally, we calculate the finite temperature phase diagrams of this two-dimensional gapless superfluid and observe two distinct critical temperatures, demonstrating the fruitful many-body effects on a paired topological superfluids. 
\end{abstract}

\maketitle

\section{Introduction}

The existence of Majorana particles has been a long-standing and important subject in theoretical high-energy physics~\cite{MF}, because they are defined by their own antiparticles and hence have many special properties compared to other elementary particles. In condensed matter physics, however, Majorana particles are referred to the quasiparticles emerging on the edge of a topological superfluid, called Majorana zero modes (MZM).  MZMs obey non-Abelian statistics and hence are expected to be the major player for a  fault-tolerant  quantum computer~\cite{MFQC,MFQC2,sarmaMajoranaZeroModes2015} . Recent rapid development on the proposal and observation of Majorana zero modes has drawn a lot of attentions in the field of condensed matter physics~\cite{MFSF}. 

To realize topological superconductors supporting Majorana zero modes, many theoretical models~\cite{Kitaev_1D,PhysRevLett.100.096407,S_Heterostructures} and experimental systems were proposed in recent years, 
including nanowires in contact with the superconductors~\cite{samuelreichSolidCaseMajorana2012}, 
quantum anomalous Hall insulator/superconductor hybrid devices~\cite{stajicLookingChiralMajoranas2020}, 
quantum spin liquids~\cite{banerjeeProximateKitaevQuantum2016a}, iron-based superconductors~\cite{wangEvidenceMajoranaBound2018}, and so on. However, the experimental evidences are still unclear and some of them face repeatable crises~\cite{kayyalhaAbsenceEvidenceChiral2020,castelvecchiEvidenceElusiveMajorana2021}.

The development of experiment techniques in ultracold atoms suggest other possible scenario to realize Majorana zero modes in the systems of topological superfluids ~\cite{MF_Cold,MZM_Fermionic_Cold_Atoms, MF_Cold_Atom_Quantum_Wires, 2D_Spin_Orbit, s_Wave_SF}.
Moreover, the quantum gases made by Rydberg-dressed atoms further open up new possibilities to study various topological phases, because the effective dipole moments, interaction strength and even the interaction range are all tubale in a wide parameter regime. In fact, many interesting many-body ground states have been predicted in different magneto-optical traps, including a two-dimensional (2D) layer, bi-layer and multi-layer systems ~\cite{huang2021twodimensional,Pikovski_2010,Zinner_2012,Babadi_2011,Baranov_2011,Cinti_2017,Potter_2010,PhysRevA.99.043624}. 

In this paper, we consider a bi-layer system with a Rydberg-dressed Fermi gas. We focus on the parameter region where the effective finite-ranged interaction are attractive in all directions. Minimizing the total energy with respect to both the amplitudes and phases of these order parameters within the self-consistent mean-field approximation, we find that the systems can host superfluid state with $s$-wave pairing, $p$-wave pairing, and a mixture of them. For the latter case, the pairing phases of these two $p$-wave in the two layers has a phase difference $\pi$, and gapless points emerge in the bulk energy spectrum. We characterize each point by calculating the quantized winding number around them and show how the zero-energy flat bands emerge at the edges spectrum. We emphasize that such a gapless topological superfluid results from a specific pairing phase locking between the
co-existing $s-$wave and $p-$wave pairing order parameters, and therefore were not considered in the literature before. We further show how these two order parameters co-exist and influence each other in the finite temperature regime. 

The article is organized as follows. In Sec. \ref{Sec: System Hamiltonian}, we introduce the Hamiltonian for a Fermi gas loaded into a 2D bi-layered system with the Rydberg-dressing interaction. In Sec. \ref{Sec:MF ground state energy}, we derive the mean-field Hamiltonian with both the amplitudes
and phases of the pairing order parameters. In Sec. \ref{Sec:Quantum Phase Diagram}, we calculate the ground state energy through variational methods and show the obtained quantum phase diagram. In Sec. \ref{Sec:Gapless Topological Superfluid}, we demonstrate that a non-trivial phase between the $s-$wave and the $p-$ wave is a topological superfluid with a finite winding number at the gapless points, leading to localized zero energy models at the edge. In Sec. \ref{Sec: Finite Temperature}, we calculate the critical temperature of the gapless topological superfluid with two order parameters, and then summarize our results in Sec. \ref{Conclusion}.

\section{Physical System and Hamiltonian}
\label{Sec: System Hamiltonian}

\subsection{Rydberg-dressed Interaction} 

\begin{figure}[htb]
\centering
\includegraphics[width=0.4\textwidth]{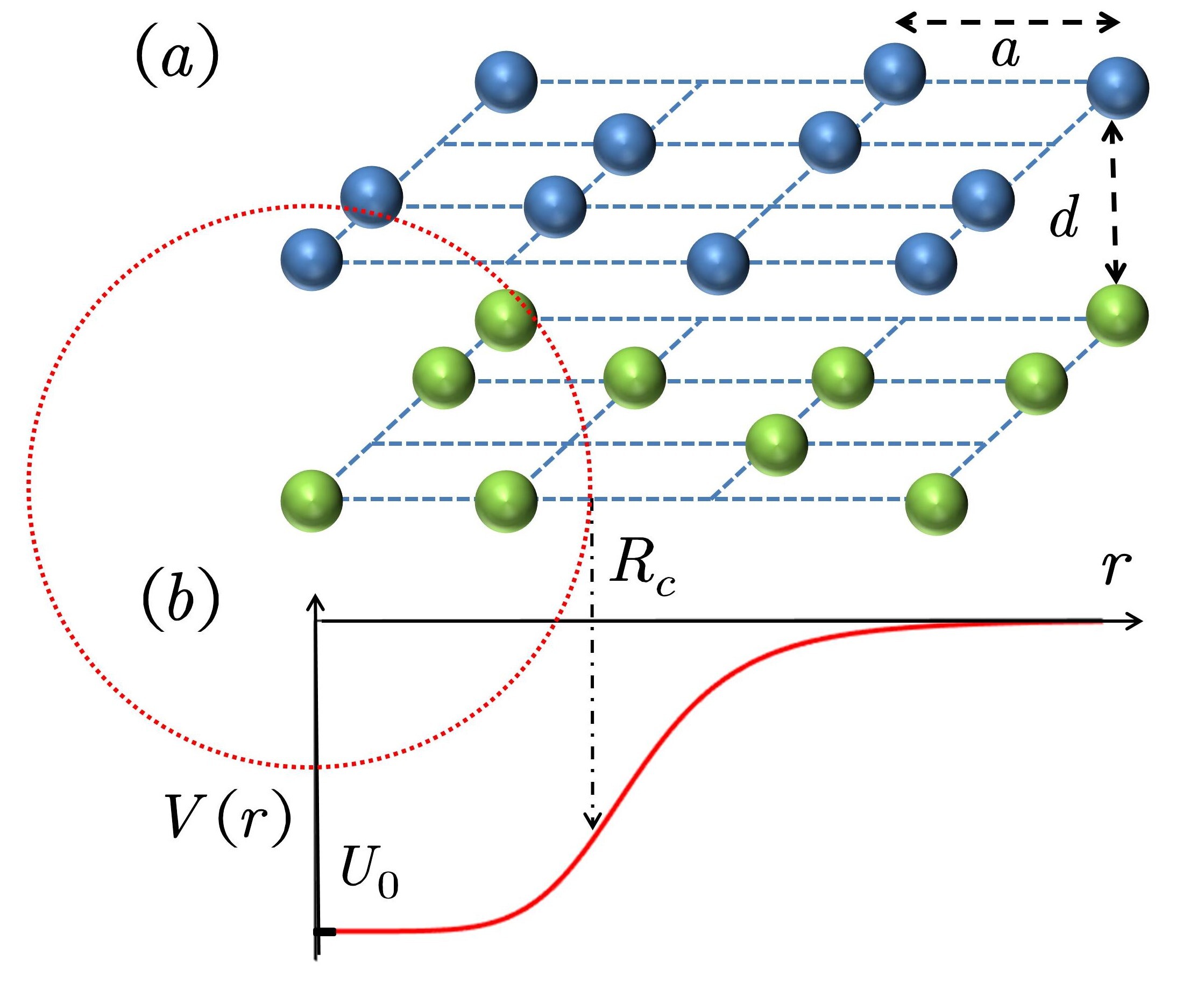}
\caption{(a) The bilayer structure with a single species Fermi gas discussed in this work.  A square optical lattice potential is applied inside the layer $(x-y)$ plane with $a$ and $d$ being the intra-layer and inter-layer lattice spacing respectively. The blue/green objects indicate fermionic Rydberg atoms loaded in the upper/lower layer. 
(b) The obtained effective Rydberg-dressed interaction, $V_{RD}( r )$ between these fermionic atoms with a soft-core strength, $U_{0}$ and a finite blockade radius $R_{c}$. See the text.  }
\label{Fig:system structure}
\end{figure}

We consider a 2D bi-layer system with a square optical lattice within the $x-y$ plane (see Fig. \ref{Fig:system structure}(a)), where a single-species Fermi gas is loaded with finite tunneling amplitudes between lattice sites in the same or opposite layers. As a result, the system could also be effectively understood as a 2D pseudo-spin $1/2$ system with an effective in-plane magnetic field via the inter-layer tunneling. In order to generate an effective interaction between these single-species fermions, an off-resonant two-photon transition is introduced to weakly couple the electronic ground state to a Rydberg excited state via an intermediate state. In the far detuning and weak coupling limit, it is known that the effective Rydberg-dressed interaction between these dressed-state atoms could be approximated by a finite ranged soft-core interaction(see Fig. \ref{Fig:system structure}(b)): $V_{\rm RD}(\mathbf{r})=\frac{U_{0}}{1+(r/R_{c})^{6}}$ within the standard perturbation and adiabatic approximation~\cite{henkelThreeDimensionalRotonExcitations2010,honerCollectiveManyBodyInteraction2010,liProbingInteractionRydbergdressed2012,plodzienRydbergDressingOnedimensional2017,pupilloStronglyCorrelatedGases2010,tongLocalBlockadeRydberg2004}. Here the effective interaction strength, $U_0$, and the interaction range, $R_c$, could be calculated explicitly from the Rabi coupling and detuning from the atomic excitations \cite{henkelThreeDimensionalRotonExcitations2010}. Many proposals discussed how to manipulate such a Rydberg-dressed interaction via various external fields~\cite{saffmanQuantumInformationRydberg2010,browaeysExperimentalInvestigationsDipole2016}.

We focus on the parameter regime to generate an effective attraction ($U_0<0$) and a finite-ranged Rydberg blockade radius ($R_c\sim a,d$, with $a$ and $d$ being the intra-layer and inter-layer lattice spacing respectively). As a result, these single-species fermions could be effectively interacting with each other with the following Hamiltonian in real space 
\begin{eqnarray}
\hat{H}&=& -t\sum_{\bfr,\sigma}\left[\hat{c}^\dagger_{\bfr,\sigma}\hat{c}^{}_{ \bfr+\hat{x},\sigma}+\hat{c}^\dagger_{ \bfr,\sigma}\hat{c}^{}_{ \bfr+\hat{y},\sigma}  +\textit{h.c.}\right]
\nonumber\\
&&-t_z\sum_{ \bfr,\sigma} \hat{c}^\dagger_{ \bfr,\sigma}\hat{c}^{}_{ \bfr,-\sigma}  
-\mu\sum_{\bfr,\sigma} \hat{c}^\dagger_{\bfr,\sigma}\hat{c}^{}_{\bfr,\sigma } \nonumber
\\
&&-\frac{1}{2} \sum_{ \bfr, \bfr',\sigma} V_{\parallel}(\bfr-\bfr')  \hat{c}^{\dagger}_{ \bfr,\sigma}\hat{c}^{\dagger}_{ \bfr',\sigma}\hat{c}^{}_{ r',\sigma} \hat{c}^{}_{ r,\sigma}
\nonumber\\
&&-\frac{1}{2} \sum_{ \bfr, \bfr',\sigma}
V_{\perp}(\bfr-\bfr') \hat{c}^{\dagger}_{ \bfr,\sigma}\hat{c}^{\dagger}_{ \bfr',-\sigma}\hat{c}^{}_{ \bfr',-\sigma} \hat{c}^{}_{ \bfr,\sigma}
\label{Eq:H_original}
\end{eqnarray}
where $\hat{c}_{\bfr,\sigma}^{\dagger}$($\hat{c}_{\bfr,\sigma}^{}$) is the creation(annihilation) operator of fermions with the layer index $\sigma=(\uparrow ,\downarrow)$, $\mu$ is the chemical potential,  and $\bfr=\left(i_{x}, i_{y}\right)$ is the in-plane coordinate. The intra-layer and inter-layer tunnelling amplitude are $t$ and $t_{z}$ respectively. Both tunneling amplitudes can be tuned independently by the lattice strength and lattice spacing. Here $V_{\perp}(\bfr-\bfr')$ and $V_{||}(\bfr-\bfr')$ are the interlayer and intra-layer interaction provided by the Rydberg-dressed interaction, $V_{RD}(\bfr)$, and defined positively for the convenience of latter discussion.

\subsection{BCS mean-field Approximation}

To investigate the topological features, we further apply Fourier transform of the field operator, $\hat{c}_{{\bfr},\sigma}=\frac{1}{\Omega}\sum_{\bfk} \hat{c}_{\bfk,\sigma} e^{i\bfk\cdot\bfr}$, to transform the Hamiltonian into the momentum space and apply the BCS mean-field approximation between pairs of fermions. Here $\Omega$ is the system area. It is well-known that the resulting mean-field Hamiltonian ($\hat{H}_{MF}$) can be written to be a $4\times 4$ matrix form within the Nambu spinor representation, $\hat{\Psi}_\bfk\equiv \left[\hat{c}_{\mathbf{k} \uparrow}, \hat{c}_{\mathbf{k} \downarrow}, \hat{c}_{-\mathbf{k} \uparrow}^{\dagger}, \hat{c}_{-\mathbf{k} \downarrow}^{\dagger}\right]^T$, and the final results can be divided in two parts:	$\hat{H}_{MF}=\hat{H}_{BCS}+E_{\mathrm{c}}$, where 
\begin{eqnarray}
\hat{H}_{BCS} &=&\frac{1}{2}\sum_{\mathbf{k}}
\hat{\Psi}_\bfk^\dagger
\left[\begin{array}{cccc}{\varepsilon_{k}} & {-t_{z}} & { {2\Delta}_{p \uparrow }^{*}} & {\Delta_{s}^{*}} \\ {-t_{z}} & {\varepsilon_{k}} & {-\Delta_{s}^{*}} & { {2\Delta}_{p\downarrow}^{*}} \\ { {2\Delta}_{p \uparrow }} & {-\Delta_{s}} & {-\varepsilon_{k}} & {t_{z}} \\ {\Delta_{s}} & { {2\Delta}_{p \downarrow}} & {t_{z}} & {-\varepsilon_{k}}\end{array}\right]\hat{\Psi}_\bfk
\label{H_BCS}
\end{eqnarray}
and the constant energy term is given by 
\begin{eqnarray}
E_{c } &=& \frac{1}{\Omega}\sum_{\bfk,\bfk^{\prime}} V_{\bfk, \bfk^{\prime}}^{\perp}\langle \hat{c}_{\bfk^{\prime}, \uparrow}^{\dagger} \hat{c}_{-\bfk^{\prime}, \downarrow}^{\dagger}\rangle\left\langle \hat{c}_{-\bfk, \downarrow} \hat{c}_{\bfk,\uparrow}\right\rangle   
	\nonumber \\
&& +\frac{1}{\Omega}\sum_{\bfk,\bfk'} V_{\bfk,\bfk^{\prime}}^{\|}\langle \hat{c}_{\bfk^{\prime}, \uparrow}^{\dagger} \hat{c}_{-\bfk^{\prime}, \uparrow}^{\dagger}\rangle\left\langle \hat{c}_{-\bfk \uparrow} \hat{c}_{\bfk,\uparrow}\right\rangle 
  \nonumber \\
&& +\frac{1}{\Omega}\sum_{\bfk,\bfk^{\prime}} V_{\bfk, \bfk^{\prime}}^{\|}\langle \hat{c}_{\bfk^{\prime}, \downarrow}^{\dagger} \hat{c}_{-\bfk^{\prime}, \downarrow}^{\dagger}\rangle\left\langle \hat{c}_{-\bfk, \downarrow} \hat{c}_{\bfk,\downarrow}\right\rangle
  \nonumber \\
&&+2 \sum_{\mathbf{k}} \varepsilon_{\mathbf{k}}
\label{Eq:E_c}
\end{eqnarray}
with $\langle \cdots\rangle$ being an expectation value and $\varepsilon_{\bfk}= -2 t( \cos k_{x}+\cos k_{y})-\mu$ being the bare kinetic energy. For simplicity, we use the layer index as the psudo-spin index and define the inter-layer pairing order parameters to be the "$s$-wave" pairing ($\Delta_s$) and the intra-layer pairing order to be the "$p$-wave" pairing ($\Delta_p$). These order parameters could be calculated through the inter- and intra-layer interaction strength as following:  
\begin{eqnarray}
\Delta_{s}(\bfk) &=&\frac{1}{\Omega}\sum_{\bfk^{\prime}} V_{\bfk, \bfk^{\prime}}^{\perp}\left\langle \hat{c}_{-\bfk',\downarrow} \hat{c}_{\bfk', \uparrow}\right\rangle
\label{Eq:Delta_s}
\\
\Delta_{p,\sigma}(\bfk) &=& \frac{1}{\Omega} \sum_{\bfk^{\prime}} V_{\bfk,\bfk^{\prime}}^{\|}\left\langle \hat{c}_{-\bfk', \sigma} \hat{c}_{\bfk',\sigma}\right\rangle
\label{Eq:Delta_p}
\end{eqnarray}
for $\sigma=\pm=\uparrow/\downarrow$.
Here $V_{\bfk,\bfk^{\prime}}^{\|,\perp}$ is the Fourier transform of the intra-layer and the inter-layer interaction matrix elements $V_{\|,\perp}(\bfr-\bfr')$ respectively.

Note that, the inter-layer pairing could be in principle either $s$-wave (singlet) or $p$-wave (triplet). But the singlet pairing must be energetically more favorable, because its symmetric orbital wavefunction always lowered the attractive inter-layer interaction energy, similar to the systems of polar molecules with an electric field perpendicular to the layer plane ~\cite{PhysRevA.96.061602,Zinner_2012,Pikovski_2010}. On the other hand, the intra-layer pairing in the $x-y$ plane must be triplet since the orbital wavefunction in the momentum space must be anti-symmetric as shown in Eq. (\ref{Eq:Delta_p}).\\

\section{mean-field Ground State Energy}
\label{Sec:MF ground state energy}
\subsection{Ground State Energy}

In order to calculate the ground state energy within the mean-field approximation, we first diagonalize the BCS mean-field Hamiltonian (Eq. (\ref{H_BCS})), and obtain the low energy excitations for the Bogoliubov quasi-particles: 
\begin{eqnarray}
\hat{H}_{BCS}&=&\frac{1}{2}\sum_{\mathbf{k}}\hat{\Gamma}_\bfk^\dagger
\left[\begin{array}{cccc}{E_{1, \mathbf{k}}} & {0} & {0} & {0} \\ {0} & {E_{2, \mathbf{k}}} & {0} & {0} \\ {0} & {0} & {E_{3, \mathbf{k}}} & {0} \\ {0} & {0} & {0} & {E_{4, \mathbf{k}}}\end{array}\right]
\hat{\Gamma}_\bfk
\label{Eq:H_BCS}
\end{eqnarray}
where $\hat{\Gamma}_\bfk\equiv \left[\hat{\gamma}_{1,\mathbf{k}}, \hat{\gamma}_{2,\mathbf{k}}, \hat{\gamma}_{3,-\mathbf{k}}^\dagger, \hat{\gamma}_{4,-\mathbf{k}}^\dagger\right]^T$ is the Nambu spinor in Bogoliubov eigemode basis with $E_{j,\bfk}$ ($j=1,\cdots 4$) being their  excitation energies. Their analytic forms may not be available for arbitrary $\Delta_s$ and $\Delta_p$, and therefore will be evaluated numerically in the rest of this paper.

Using the anti-commutation relationship between fermion operator, $\hat{\gamma}_{j,\bfk}$, and $\langle \hat{\gamma}_{j,\bfk}^\dagger \hat{\gamma}_{j,\bfk}\rangle =0$ for the ground state expectation value, we could easily derive the mean-field ground state energy to be (combined with the constant term, $E_c$ in Eq. (\ref{Eq:E_c}))
\begin{widetext}
\begin{eqnarray}
E_{G}&=&E_C+\frac{1}{2} \sum_{j,\bfk} E_{j,\bfk}
=\frac{1}{\Omega}\sum_{\bfk,\bfk^{\prime}} V_{\bfk, \bfk^{\prime}}^{\perp}\langle \hat{c}_{\bfk^{\prime}, \uparrow}^{\dagger} \hat{c}_{-\bfk^{\prime}, \downarrow}^{\dagger}\rangle\left\langle \hat{c}_{-\bfk, \downarrow} \hat{c}_{\bfk,\uparrow}\right\rangle   
+\frac{1}{\Omega}\sum_{\bfk,\bfk'} V_{\bfk,\bfk^{\prime}}^{\|}\langle \hat{c}_{\bfk^{\prime}, \uparrow}^{\dagger} \hat{c}_{-\bfk^{\prime}, \uparrow}^{\dagger}\rangle\left\langle \hat{c}_{-\bfk \uparrow} \hat{c}_{\bfk,\uparrow}\right\rangle 
  \nonumber\\
  && +\frac{1}{\Omega}\sum_{\bfk,\bfk^{\prime}} V_{\bfk, \bfk^{\prime}}^{\|}\langle \hat{c}_{\bfk^{\prime}, \downarrow}^{\dagger} \hat{c}_{-\bfk^{\prime}, \downarrow}^{\dagger}\rangle\left\langle \hat{c}_{-\bfk, \downarrow} \hat{c}_{\bfk,\downarrow}\right\rangle
 +2\sum_{\mathbf{k}}\varepsilon_{\mathbf{k}}
  +\frac{1}{2}\sum_{j,\bfk} E_{j,\bfk}
\label{Eq:E_G}
\end{eqnarray}

From the expression above, one could see that the ground state energy depends on the ground state expectation value of the pairing operators, $\langle \hat{c}_{-\bfk,\sigma}\hat{c}_{\bfk,\sigma'}\rangle$, which also appears in the definition of the order parameter, $\Delta_{s}(\bfk)$ and $\Delta_{p,\sigma}(\bfk)$, see Eqs. (\ref{Eq:Delta_s}) and (\ref{Eq:Delta_p}). In this paper, we will use variational method to calculate the ground state energy by parametrizing these order parameters according to the lattice symmetry. More precisely, we could expand the intra- and inter-layer interaction as following:
\begin{eqnarray}
V^{\|,\perp}_{\mathbf{k},\mathbf{k}^{\prime}} &=& \sum_{mn\in Z}V_{mn}^{\|,\perp}
\cos\left[m(k_x-k_x')a+n(k_y-k_y')a\right]
\\
\Delta_{s}(\bfk)&=&
\sum_{m,n\in Z}\Delta^{s}_{mn}
\left[\cos\left(mk_xa+nk_ya\right)
+\cos\left(nk_xa-mk_ya\right)\right]
\\
\Delta_{p,\sigma}(\bfk)&=&
\sum_{m,n\in Z}\Delta^{p,\sigma}_{mn}
\left[\sin\left(mk_xa+nk_ya\right)
-i\sin\left(nk_xa-mk_ya\right)\right]
\\
\langle \hat{c}_{-\bfk,\uparrow}\hat{c}_{\bfk,\downarrow}\rangle
&=&\sum_{m,n\in Z}c^s_{mn}
\left[\cos\left(mk_xa+nk_ya\right)
+\cos\left(nk_xa-mk_ya\right)\right]
\\
\langle \hat{c}_{-\bfk,\sigma}\hat{c}_{\bfk,\sigma}\rangle
&=&
\sum_{m,n\in Z}c^{p,\sigma}_{mn}
\left[\sin\left(mk_xa+nk_ya\right)
-i\sin\left(nk_xa-mk_ya\right)\right]
\end{eqnarray}
\end{widetext}

Here $\Delta^s_{mn}$, $\Delta^{p,\sigma}_{mn}$, $c^s_{mn}$ and $c^{p,\sigma}_{mn}$ are all complex numbers, which could be determined later through a variational approach. $m$ and $n$ are integers for the index used in the reciprocal lattice. In the literature before, it is usually assumed that the pairing order parameters are proportional to the interaction strengths for simplicity, but this assumption  cannot be justified when considering the competition between order parameters. In order to consider these order parameters correctly, we note that $\Delta^s_{mn}$ and $\Delta^{p,\sigma}_{mn}$ could be calculated from $c^s_{mn}$ and $c^{p,\sigma}_{mn}$ through the definition of gap function in Eqs. (\ref{Eq:Delta_s}) and (\ref{Eq:Delta_p}). We will express their relationship in more details below for variational methods.

\begin{figure}[htb]
\centering  
\includegraphics[width=0.45\textwidth]{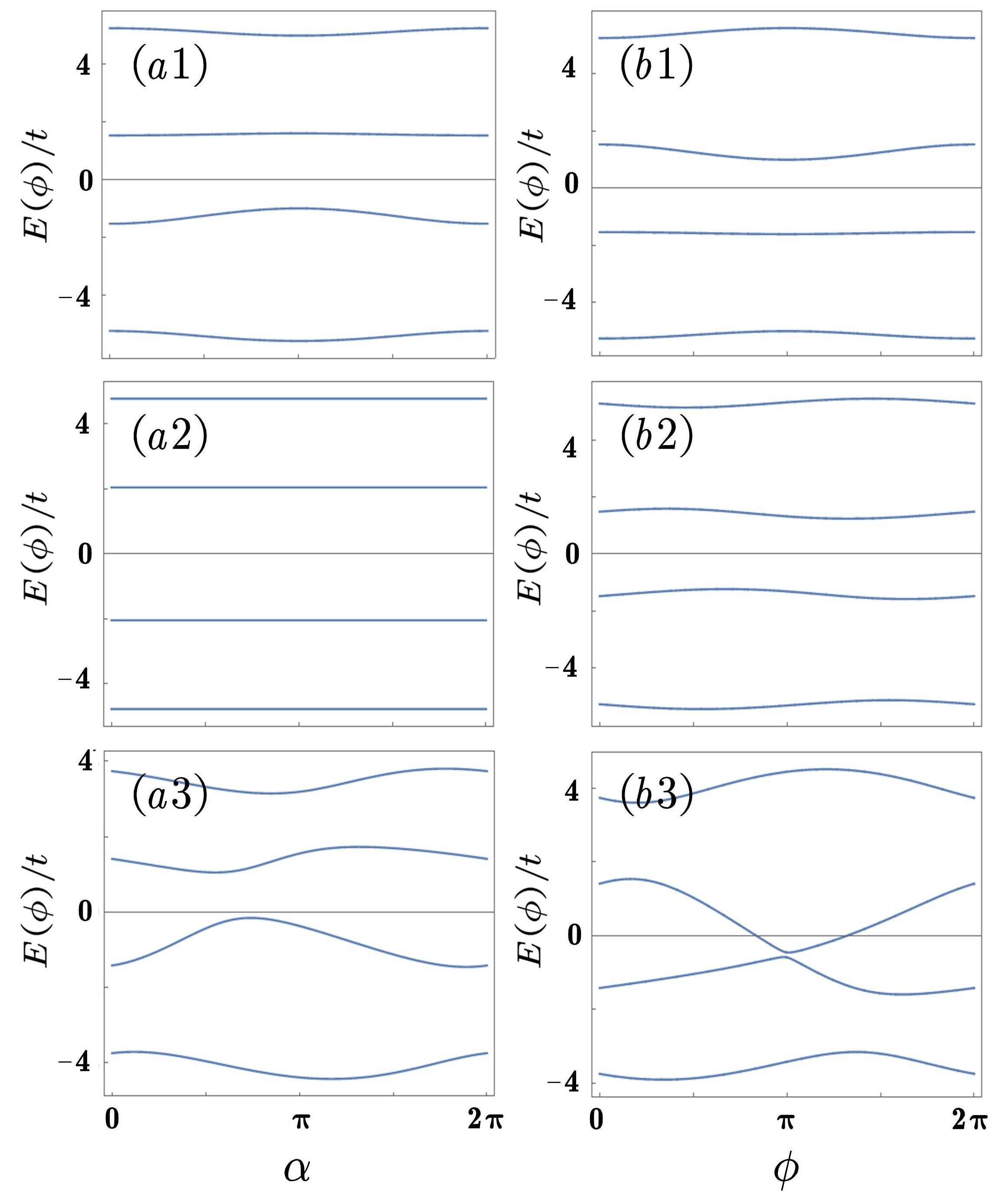}
\caption{
(a1)-(a3) are eigenstate energies $E_j$ of the BCS Hamiltonian in Eq. (\ref{Eq:H_BCS}) as a function of the pairing phase, $\alpha=\alpha_\uparrow-\alpha_\downarrow$.
(b1)-(b3) show the same results as a function of another pairing phase, $\phi=\phi_\uparrow-\phi_\downarrow$. Here $\left(k_{x}, k_{y}\right)=(0,\pi/2)$,$ (\pi/2,0)$, and $(\pi/2,\pi/2)$ for the upper, middle and lower panels respectively. Other parameters are $\Delta_{0}^{s} / t=\Delta_{1}^{p} / t=1, t_{z} / t=2$, and $\mu / t=1$. 
}
\label{Fig:Eigenenergy_spectrum}
\end{figure}

\subsection{Variational Approach and Pairing Phase Dependence of the Ground State Energy}

Although the expression of order parameters above could be applied to a general finite ranged-ranged interaction, it is still much more intuitive and transparent to start with a finite range interaction range, i.e. $R_c\sim a, d$,   with $a$ and $d$ being the intra-layer and inter-layer lattice spacing respectively. For such a finite ranged attractive interaction, it is reasonable for us to first consider the nearest neighboring terms of pairing only, i.e. $|m|,|n|,|m\pm n|\leq 1$, and neglect the longer-ranged pairing. 
Since our work is to emphasize the effects of these pairing phases on the ground state properties, it is much more instructive to concentrate on the coupling between the nearest neighboring intra-layer and inter-layer pairing only. Including longer-ranged pairing requires much more pairing phases to be calculated self-consistently, and makes it difficult to interpret the mechanism. We will investigate the pairing phase dependence for this long-ranged interaction in the future.
As a result, we could simplify the general expression of the gap functions in the last section and obtain,
\begin{eqnarray}
\Delta_{s}(\bfk)&=&
\Delta^s_{0}+\Delta^s_{1}\left[\cos (k_xa)+\cos (k_ya)\right]
\\
\Delta_{p,\sigma}(\bfk)&=&
2ie^{i\phi_\sigma}\Delta^{p}_{1}
\left[\sin\left({k}_xa\right)
+ie^{i\alpha_\sigma}\sin\left({k}_ya\right)\right] \label{Eq:Gphase}
\\
\langle \hat{c}_{-\bfk,\uparrow}\hat{c}_{\bfk,\downarrow}\rangle
&=&c^s_{0}+c^s_{1}\left[\cos(k_xa)+\cos(k_ya)\right]
\\
\langle \hat{c}_{-\bfk,\sigma}\hat{c}_{\bfk,\sigma}\rangle &=&
2ic^{p}_{1}
\left[\sin\left(k_xa\right)
+i\sin\left(k_ya\right)\right].
\end{eqnarray}
Here we have set the pairing phase of the $s$-wave pairing order parameter to be zero, and separate the phases of theses $p$-wave order parameters, so that $\Delta^s_{0/1}$,  $\Delta^p_{1}$, $c_{0/1}^{s}$ and $c_{1}^{p}$ are all defined to be positive. $\alpha_\sigma$ and $\phi_\sigma$ are the four relative phases for the pairing order parameters as indicated above. Extension to a longer interaction range is straightforward but will not be considered in this paper.

Using the expression above, it is easy to directly connect the relationship between $\Delta^{s,p}_{0,1}$ and $c^{s,p}_{0,1}$ through the definition of order parameters in Eqs. (\ref{Eq:Delta_s}) and (\ref{Eq:Delta_p}). We have
\begin{eqnarray}
\Delta_{0}^s &= V_{0}^{\perp}c_{0}^{s}
\\
\Delta_{1}^s &=\frac{1}{2}V_{1}^{\perp}c_{1}^{s}
\\
\Delta_{1}^p &=\frac{1}{2}V_{1}^{\|}c_{1}^{p}
\end{eqnarray}
and therefore the mean-field ground state energy in Eq. (\ref{Eq:E_G}) can be calculated directly to be 
\begin{eqnarray}
E_G=\sum_{\mathbf{k}}( 2\xi_{\mathbf{k}}-\frac{1}{2} \sum_{j} E_{j,\bfk} )+\frac{|\Delta_{0}^{s}|^2}{V_{0}^{\perp}} +\frac{2|\Delta_{1}^{s}|^2}{V_{1}^{\perp}}+\frac{2 |\Delta_{1}^{p}|^{2}}{V_{1}^{\|}}
\nonumber
\end{eqnarray}
Here the three order parameters, $\Delta^s_{0,1}$ and $\Delta^p_1$, are then treated as independent variational parameters.

However, besides the magnitude of order parameters, the ground state energy also depends on the relative values between these pairing phases, $(\alpha_{\uparrow},\alpha_{\downarrow},\phi_{\uparrow},\phi_{\downarrow})$. They are embedded inside the expression of the Bogoliubov eigenstate energy, $E_{j,\bfk}$. We have examined all the possible combinations of these pairing phases, but present results only using their relative values, $\alpha \equiv \alpha_{\uparrow}-\alpha_{\downarrow}$ and $\phi \equiv \phi_{\uparrow}-\phi_{\downarrow}$ by setting $\alpha_{\downarrow}=\phi_{\downarrow}=0$ in the rest of this paper. It is because this could give the most representative results without missing other information. As an example, we could analytically calculate the eigenvalues of Bogoliubov excitations for $(\alpha,\phi)=(0,0)$ to be
\begin{widetext}
\begin{eqnarray}
E_{j, \mathbf{k} }=\pm\left[\varepsilon_{ \mathbf{k} }^{2}+t_{z}^{2}+4\left|\Delta_{1}^{p}\right|^{2}\left(\sin ^{2} \left(k_{x} a\right)+\sin ^{2} \left(k_{y} a\right)\right)+\left|\Delta_{0}^{s}\right|^{2} \pm 2  \sqrt{\varepsilon_{ \mathbf{k}}^{2}t_{z}^{2}+t_{z}^{2}\left|\Delta_{0}^{s}\right|^{2}+4\left|\Delta_{1}^{p}\right|^{2}\left|\Delta_{0}^{s}\right|^{2}  \sin ^{2} \left(k_{x} a\right)}\right]^{1 / 2}
\label{Eq:E_jk_00}
\end{eqnarray}
For $(\alpha,\phi)=(\pi,\pi)$, we have
\begin{eqnarray} 
E_{j, \mathbf{k} }=\pm\left[\varepsilon_{ \mathbf{k} }^{2}+t_{z}^{2}+4\left|\Delta_{1}^{p}\right|^{2}\left(\sin ^{2} \left(k_{x} a\right)+\sin ^{2} \left(k_{y} a\right)\right)+\left|\Delta_{0}^{s}\right|^{2} \pm 2\left|t_{z}\right| \sqrt{\varepsilon_{ \mathbf{k}}^{2}+\left|\Delta_{0}^{s}\right|^{2}+4\left|\Delta_{1}^{p}\right|^{2} \sin ^{2} \left(k_{x} a\right)}\right]^{1 / 2}
\label{Eq:E_jk_pipi}
\end{eqnarray}
\end{widetext}
Analytical forms of these eigenstate energies could be also obtained in other specific values of $(\alpha,\phi)$.
However, for a general value of $(\alpha,\phi)$, we will evaluate it numerically for the variational calculation below. In Fig. \ref{Fig:Eigenenergy_spectrum}(a1)-(a3) and (b1)-(b3), we show how the numerically calculated eigenenergies changes as a function of these pairing phases.

\subsection{The system symmetry}

From the Anderson pseudo-spin point of view~\cite{Anderson1958}, the $SU(2)$ and $Z_2$ symmetries are broken explicitly in the BCS mean-field Hamiltonian (see Eq. (\ref{H_BCS})). Nevertheless, we find that, by selecting appropriate parameters, say $t_{z}=0$, $\phi_{\uparrow}=-\phi_{\downarrow}$, and $\alpha_{\uparrow}=\pi+\alpha_{\downarrow}$, the bilayer system we study here becomes time-reversal symmetric. More precisely speaking, the chiral directions of the two superfluids in the upper and lower layers are opposite, making the time-reversal symmetry restored if reversing the parity in the $z$-axis. However, if the inter-layer tunneling is finite, such time-reversal symmetry disappears as expected, both in the original Haimitonian and the mean-field one. In terms of the mean-field Hamiltonian, the pairing term in low energy limit can be expressed as $\Delta(k) = i \sigma_2 (\Delta_s \sigma_0 + \Delta_p (k_x \sigma_1 + i e^{i\alpha} k_y \sigma_2 ))$. The value of $\alpha$ determines the time-reversal symmetry of the system. We find that in the self-consistent mean-field treatment, $\alpha$ can be $\pi$. The time-reversal operation $T:= i \sigma_2 K$ maps $\Delta(k)$ to $\Delta(-k)$ indicates the preserving of the time-reversal symmetry. In our self-consistent mean-field treatment, we do not add other constraints besides the breaking of $U(1)$ symmetry. However, since the relative values of these phases may be pinned by the interaction effects, some symmetry (say the $C_4$ rotational symmetry of original Hamiltonian in the $x-y$ plane) may be also broken in the exotic phases we discussed in the context of Fig.\ref{Fig:band structure and edge state}. \\

\section{Quantum Phase Diagram}
\label{Sec:Quantum Phase Diagram}

\subsection{Competition and Co-existence of $s$- and $p$-wave Order Parameters}

\begin{figure*}[tbp]
\centering
\includegraphics[width=0.98\textwidth]{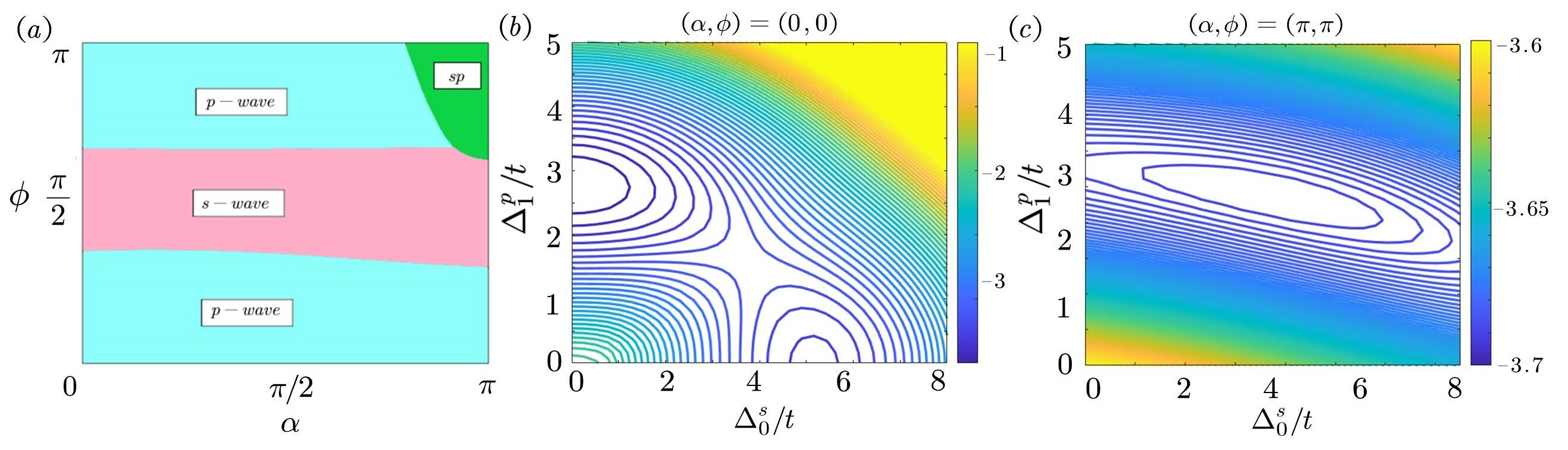}
\caption{(a)Diagrams of dominant pairing symmetry in terms of the pairing phase, $(\alpha, \phi)$, for $V_{0}^{\perp} / t=11$ , $V_{1}^{\|} / t=6.2$ and $t_z/t=0.8$, obtained by the self-consistent variational method.}  The pink/blue area are for the $s-$wave/$p-$wave pairing states, and the associated phase transitions between them are first order. The green area is the state with the coexistence of  $s-$ and $p$-waves at the same time, and its phase transition boundary with the other two phases are second order. (b) Variational energy contours for the pairing phase, $(\alpha, \phi)=(0,0)$, as a function of $\Delta^s_0/t$ and $\Delta^p_1/t$ ($\Delta^s_1=0$ for these cases). It shows two local minimum in the variational energy for the $s$-wave and $p-$wave states respectively. Their relative value determines the true ground state and hence provides a first order phase transition. (c) Same as (b) for $(\alpha, \phi)=(\pi, \pi)$. One could see that the energy minimum occurs at a point of finite values in $\Delta^s_0$ and $\Delta^p_1$. It indicates the higher order coupling between the $s-$wave and $p-$wave order parameters.
\label{Fig:Quantum Phase diagram in phase}
\end{figure*}
In Fig. \ref{Fig:Quantum Phase diagram in phase}(a), we show the calculated quantum phase diagram in terms of the two pairing phases, $\alpha$ and $\phi$. The result is obtained by a given set of system parameters, $V_0^\perp/t=11$, $V_1^\perp/t=5.8$, $V_1^\|/t=6.2$, $t_z/t=0.8$, and $\mu=0.2$. The ground state is obtained by minimizing the ground state energy in the parameter space, $(\Delta_0^s,\Delta_1^s,\Delta_1^p)$.  One can see that when the relative pairing phase between the two layers (i.e. $\alpha$) is small, the ground state is mainly $s-$wave for $\phi\sim \pi/2$ and becomes $p-$wave otherwise. The phase boundary between these two phases depend on $\alpha$ weakly. The transition between these two phases are found to be first order, as shown in Fig. \ref{Fig:Quantum Phase diagram in phase}(b). Such a result is reasonable because, when only the onsite inter-layer $s-$wave pairing is present (i.e. $\Delta_0^s\neq 0$, $\Delta_1^s=\Delta_1^p=0$), all the pairing phase dependence of the ground state energy disappears. On the other hand, when the $p-$wave order parameters dominant, their relative phase respect to the $s-$wave also becomes unimportant, while the ground state energy strongly depends on the relative phase, $\phi$, between the upper and lower layers. The lowest energy for small $\alpha$ regime stays at $\phi=0,\pi$ as expected.

Note that, from energetic point of view, one could easily expect that the $s-$wave pairing will dominate as the two layers are very close (i.e. $d\ll a$), while the $p-$wave become dominant as the two layers are far away from each other (i.e. $d\gg a$). The phase diagram here is obtained by changing the pairing phases for a fixed value of $d/a$ (here given by $t_z/t=0.8$). Such pairing phase dependence in ground state energy could lead to a certain spontaneous phase transition, since these pairing phases are automatically determined by the interaction effects. 

Besides the two traditional pairing mechanism shown above, we find that a hybrid ground state, where both $\Delta_0^s$ and $\Delta_1^p$ are finite (but within the parameter regime, $\Delta_1^s=0$). 
It appears in a regime near $(\alpha, \phi)=(\pi,\pi)$, which was not investigated before. From the self-consistent mean-field calculation demonstrated here, the phase transition between this $sp$-coexisting ground sate appears through a second order phase transition with a lower energy than either $s-$wave or $p-$wave states alone. 
The presence of such a $sp$-coexisting state is from the fact that near $(\alpha, \phi)=(\pi,\pi)$, $\Delta_0^s$ and $\Delta_1^p$ are coupled in a higher order terms in Landau's free energy expression through the complicated Bogoliubov excitation spectrum. 

\subsection{Quantum Phase Diagram in terms of the System Parameters}

In order to investigate the phase diagram in terms of the realistic system parameters, we further express the quantum phase diagram in terms of interaction strengths, $V_0^\perp$ and $V_1^\|$. The ground states are then determined by minimising the full variational energy ($E_G$) in terms of three variational order parameter, ($\Delta_0^s$, $\Delta_1^s$, $\Delta_1^p$), and two pairing phases, $(\alpha,\phi)$.  Note that, both $V_0^\perp$ and $V_1^\|$ can be directly calculated from the full effective Rydberg-dressed interaction between dressed states. More generally, we analytically express their values in terms of a general inter-layer and intra-layer interaction, i.e.
$V_0^\perp=V_{0,0}^\perp$, $V_1^\perp=V_{\pm 1,0}^\perp=V_{0,\pm 1}^\perp$, and $V_{1}^\|=V_{\pm 1,0}^\|=V_{0,\pm 1}^\|$, where
\begin{eqnarray} 
V_{m,n}^{\perp}&\equiv &\frac{U_0}{\left[d^{2}+(n a)^{2}+(ma)^{2}\right]^{3}+R_{c}^{6}}
\\
V_{m,n}^{\|} &\equiv &\frac{U_0}{[(na)^2+(ma)^2]^3+R_{c}^6}
\end{eqnarray}
with $n,m\in Z$ being the lattice site index. As a result, both $V_{1}^{\|}$ and $V_{0}^{\perp}$ can be tuned independently by changing $U_0$ and $d$, keeping the lattice constant $a$ and the interaction range $R_c$ the same. The value of inter-layer and inter-site interaction, $V_1^\perp$ is then changed accordingly. 

\begin{figure}[htb]
\centering
\includegraphics[width=0.38\textwidth]{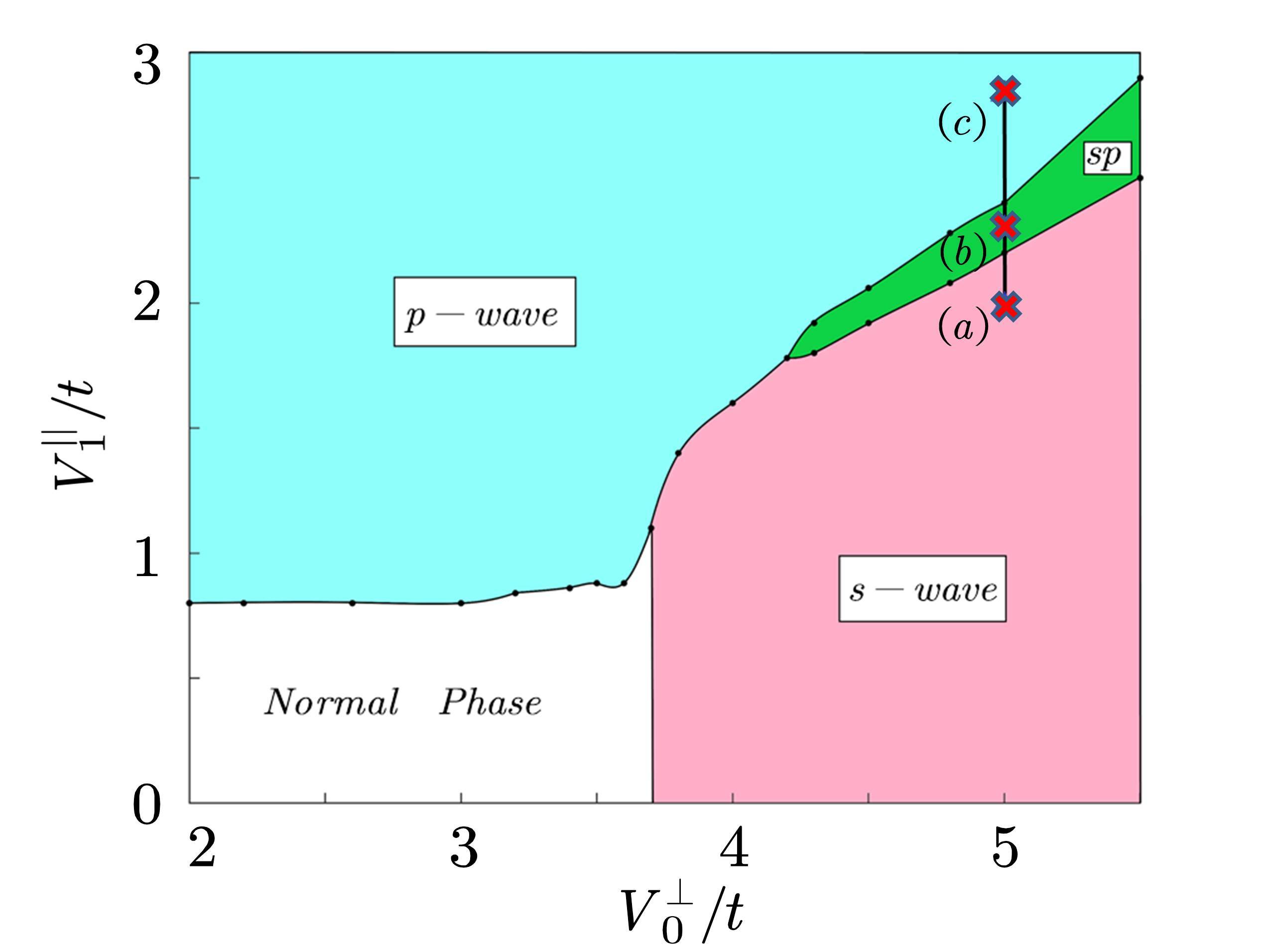}
\caption{Quantum phase diagram as a function of inter-layer and intra-layer interaction strengths, $V^{\perp}_0/t$ and $V^{\|}_1/t$. The pink and blue area are ground states of $s-$wave and $p-$wave pairing order parameters only. The green area stands for the ground state with the co-existance of these two order parameters at the same time (with $(\alpha,\phi)=(\pi,\pi)$). In the white region, the ground state is a normal state without order parameters within our variational method. This is due to the presence of a finite interlayer tunneling and the competition between the $s-$ and $p-$wave pairing state. Here we use $t_z=0.8$.
The vertical line and three cross points (labeled by (a), (b) and (c)) in the upper right corner are the points for the three finite temperature phase diagrams in Fig. \ref{Fig:Finite temperature}.
}
\label{Fig:quantum phase diagram}
\end{figure}

In Fig. \ref{Fig:quantum phase diagram}, we show the calculated quantum phase diagram in terms of $V_{1}^{\|}$ and $V_{0}^{\perp}$ for $t_z/t=0.8$ and $\mu/t=0.2$. Several interesting and important properties could be observed: First, in the limit of zero interaction strength (white regime), there is no superfluid order parameter, because the finite inter-layer tunneling plays as a magnetic field to open a gap between different pseudo-spin components, suppressing the formation of Cooper pairs. In principle, the ground state could be a FFLO state due to the Fermi wavevector mis-matching, while we did not include this order parameters in our mean-field approach to simplify the calculation. 
This simplification is justified here because the critical temperature of the FFLO state is known much lower than regular superfluids due to its  non-zero condensate momentum. (For example, the estimated $T_c$ of such a FFLO state of $Na-Li$ polar molecules in a bilayer system is just about 10$nK$~\cite{PhysRevA.96.061602}, well below the temperature current experiments.) Therefore, we believe our calculated quantum phase diagram should not be affected much even considering the FFLO state at a finite temperature.

Secondly, in the inter-mediate and stronger interaction regime, we observe the $s-$ ($p-$wave) superfluid in the regime when $V_0^\perp$($V_1^\|)$ becomes dominant in the blue(pink) regime. This reflects the fact that this bilayer system has a great flexibility to investigate the quantum phase transition between these two different superfluids via changing Rydberg-coupling strength or the inter-layer distance directly. As described above, the phase boundary between these two phases are first order without inter-mediate phase if the interaction is not strong.

However, in a stronger interaction regime, we do find the possibility to have a co-existing $s-$wave and $p-$wave superfluid, where the pairing phase becomes $(\alpha, \phi)=(\pi,\pi)$, as shown in the upper right corner of Fig. \ref{Fig:Quantum Phase diagram in phase}(a). We have to emphasize that, if one fixes the relative phases between $s-$ and $p-$wave order parameters to be zero ($\alpha=\phi=0$) as one usually did in the literature, we will not be able to find such a co-existing multi-order superfluidity from the variational approach. Although the regime of such an $sp-$coexisting phase seems not very large, it still provides an important route to investigate a new topological superfluidity, i.e. the traditional chiral $p-$wave superfluids could be coupled and paired together, making a possible new topological superfluid, which we will study in more details later. 

\section{Gapless Topological Superfluid}
\label{Sec:Gapless Topological Superfluid}
\subsection{Band Structure and Edge States}

\begin{figure}[htb]
\flushleft 
\includegraphics[width=0.5\textwidth]{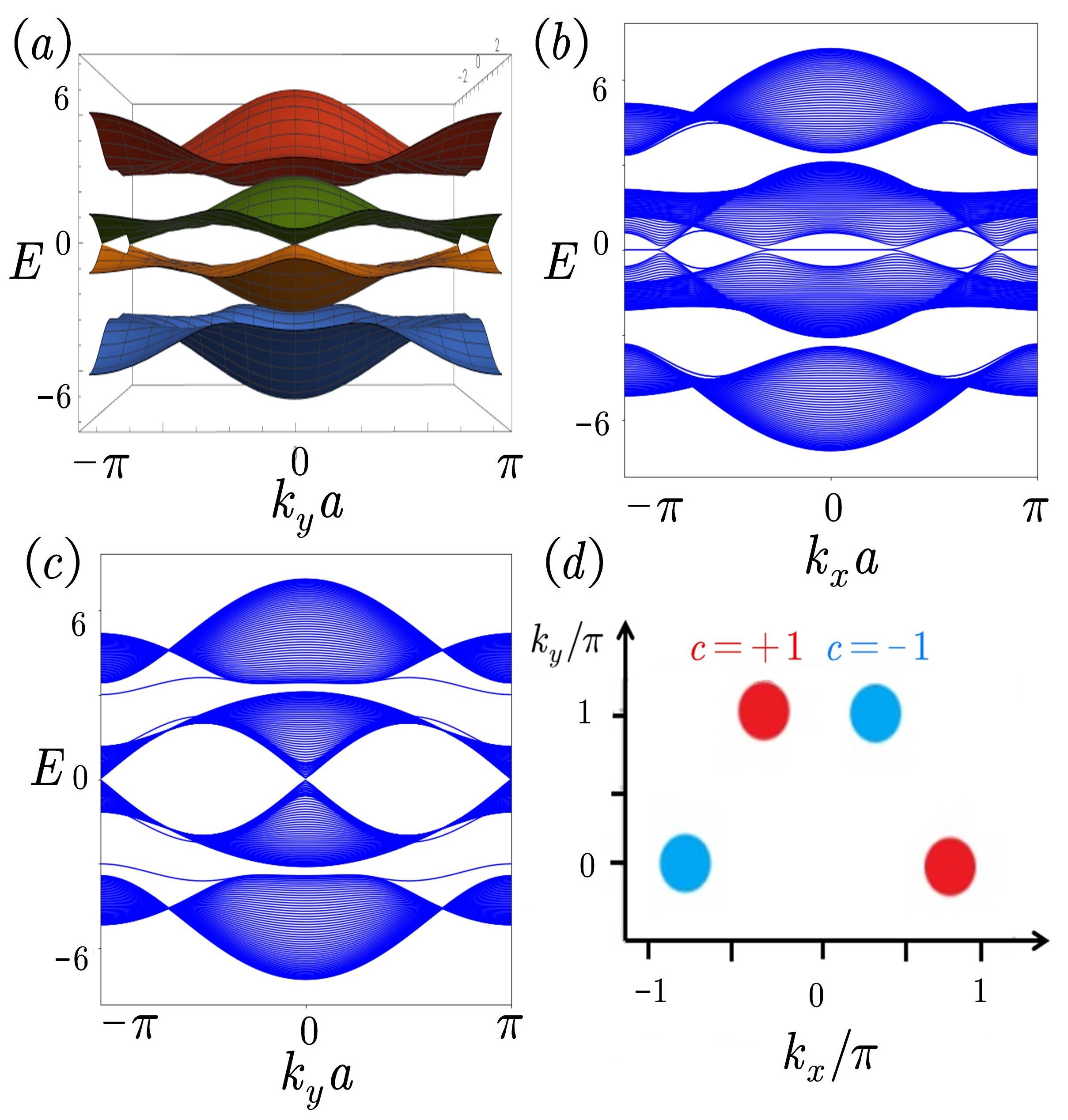}
\caption{(a) Band Structure of the the gapless topological superfluid calculated in a periodic boundary condition for $L_{x}=L_y=100$. Here we use $\Delta_{0}^{s} / t=\Delta_{1}^{p} / t=1$, $t_{z} / t=2$, and $\mu / t=1$. (b) Same as (a) but along the $k_{x}$ direction with an open boundary condition in the $y$ axis. (c) Same as (a) but along the $k_{y}$ direction with an open boundary condition in the $x$ axis.
(d) shows the position of gapless points for the gapless topological superfluid. The gap-closing points with red/blue color indicate positive/negative charges for $W=+1/-1$. 
}
\label{Fig:band structure and edge state}
\end{figure}

Since we are more interested in the phase with the co-existence of both $s-$ and $p-$wave superfluids, it will be more instructive to investigate its band structure and edge state properties first. In Fig. \ref{Fig:band structure and edge state}(a) we show the calculated band structure for $L_x=L_y=100$ with a periodic boundary condition in both $x$ and $y$ directions. We have set $(\alpha, \phi)=(\pi,\pi)$  and find several gapless points in the band structure. Note that such gapless structure does not exist if the relative phases between these order parameters are the same (i.e. $\alpha=\phi=0$). In other words, the pairing phase modulation makes two gapped superfluids ($s-$ and $p-$waves) co-exist and close their gaps, leading to interesting topological superfluid similar to nodal superconductors~\cite{Matsuura_2013, PhysRevB.73.214502, PhysRevB.83.224511, PhysRevLett.105.217001, PhysRevLett.105.097002} as shown below. 

To investigate if there could be any edge states due to the topological properties in the bulk, in Figs. \ref{Fig:band structure and edge state}(b) and (c), we show the calculated band structures with an open boundary condition in the $y$ and $x$ directions (i.e., $k_x$ and $k_y$ are good quantum numbers respectively). We could find that their band structures are very different due to the phase locking between the inter-layer pairing and the intra-layer pairing order parameters. More importantly, we could clearly find the localized state band energy in both sides. This reflects the bulk-edge corresponding principle of the topological matter. When the boundary is open in the $y$ direction, the localized edge mode has a flat band, connecting different gapless points projected on the edge Brillouin zone, as shown in Fig. \ref{Fig:band structure and edge state}(b). As we will show below, this indicates the quantized topological charge for these gapless points.

\subsection{Topological Charge of Gapless Points}

In order to investigate the topological properties of this new phase in more details, here we calculate the possible topological charge of these gapless points, whose positions, $(k^0_x,k^0_y)$, could be calculated easily from the eigenmode spectrum, see Eq. (\ref{Eq:E_jk_pipi}). The gap is closed at $k^0_{y}=0,\pi$, and
$\sqrt{\varepsilon_{\bfk^0}^{2}+\left|\Delta_{0}^{s}\right|^{2}+4\left|\Delta_{1}^{p}\right|^{2} \sin ^{2} (k_{x}^0a)} \pm t_{z}=0$, leading to four solutions distributed in the momentum space as shown in Fig. \ref{Fig:band structure and edge state}(d). The red/blue colors indicate positive/negative topological charges as we will show below.

It is well-known \cite{PhysRevB.89.235127,PhysRevB.97.155134,PhysRevB.83.205101}  that in three dimensions, the topological charge is defined to be the surface integral of Berry curvature around a topological monopole, i.e. the Chern number.
In two dimensionals, a loop that encircles the gapless point defines the topological charge, which is then equivalent to the winding number.
There are several methods to calculate the topological charges around the gapless points. Here we note that the band disperses linearly in the momentum space through the gapless point, so that we could obtain the effective Hamiltonian around  $\left(k_{x}^{0}, k_{y}^{0}\right)$ up to the linear terms, i.e. $\mathbf{k}=\mathbf{k}^{0}+\mathbf{q}+{\cal O}(q^2)$, and hene $\hat{H} (\mathbf{k} )= \hat{H}_{0}(\mathbf{k}^{0})+ \hat{H}_{1}(\mathbf{q} )+{\cal O}(\mathbf{q}^2)$. Here zeroth order term of the Hamiltonian is
\begin{eqnarray}
H_{0}\left( \mathbf{k} ^{0}\right)=\varepsilon_{ k }\left(\sigma_{3} \otimes \sigma_{0}\right)-\Delta_{s}\left(\sigma_{2} \otimes \sigma_{2}\right)-t_{z}\left(\sigma_{3} \otimes \sigma_{1}\right) \nonumber \\
\quad-2 \Delta_{p}\left[\sin \left(k_{x}^{0}a\right)\right]\left(\sigma_{2} \otimes \sigma_{0}\right)-2 \Delta_{p}\left[\sin \left(k_{y}^{0}a\right)\right]\left(\sigma_{1} \otimes \sigma_{3}\right) \notag \\
\end{eqnarray}
and the leading order Hamiltonian is
\begin{eqnarray}
\hat{H}_{1}(\mathbf{q})&=&2 t [q_{x} \sin \left(k_{x}^{0}a\right)+q_{y} \sin \left(k_{y}^{0}a\right)]\left(\sigma_{3} \otimes \sigma_{0}\right)
\nonumber\\
&&-2 \Delta_{1}^{p} q_{x} \cos \left(k_{x}^{0}a\right)\left(\sigma_{2} \otimes \sigma_{0}\right)
\end{eqnarray}
Note that we use the direct product of two Pauli matrix for a general expression of the original $4\times 4$ matrix.

In order to get the effective two-band Hamiltonian, we first numerically solve the $4 \times 4$ unperturbed Hamiltonian, $\hat{H}_{0}(\bfk^{0})$, and express the leading order term, $\hat{H}_{1}(q)$, in this eigenstate basis. We then obtain the $2\times 2$ effective Hamiltonian by  projecting the original Hamiltonian to two bands near zero energy.
\begin{eqnarray}
H _{ eff }(\mathbf{q} )&=&\left[\begin{array}{cc}
\left\langle v_{1}\left|H_{1}(\mathbf{q})\right| v_{1}\right\rangle & \left\langle v_{1}\left|H_{1}(\mathbf{q})\right| v_{2}\right\rangle \nonumber \\
\left\langle v_{2}\left|H_{1}(\mathbf{q})\right| v_{1}\right\rangle & \left\langle v_{2}\left|H_{1}(\mathbf{q})\right| v_{2}\right\rangle
\end{array}\right]
\\&=& v_{x}\left( \mathbf{k} ^{0}\right) q_{x} \sigma_{x}^{\prime}+v_{y}\left( \mathbf{k} ^{0}\right) q_{y} \sigma_{y}^{\prime}.
\end{eqnarray}
In the last line, we have transformed the $2 \times 2$ matrix into a new basis, expressed by $\{\sigma_{x}^{\prime}$ and $\sigma_{y}^{\prime}\}$ in order to match the general form of gapless points with coefficients $v_{x}\left( \mathbf{k} ^{0}\right)$ and $v_{y}\left( \mathbf{k} ^{0}\right)$ respectively. In order to calculate the winding number around these gapless points more conveniently, we expresses the final effective Hamiltonian in polar coordinates by $v_{x}\left( \mathbf{k} ^{0}\right) q_{x}=R(\mathbf{q}) \sin \theta_\bfq \cos \phi_\bfq$, $v_{y}\left( \mathbf{k} ^{0}\right) q_{y} =R(\bfq) \sin \theta_\bfq \sin \phi_\bfq$, so that  
\begin{eqnarray}
H_{ eff }( \mathbf{q} )=R(\mathbf{q} )\left(\begin{array}{cc}
0 & e^{i \theta_{ \mathbf{q} }} \\
e^{-i \theta_{ \mathbf{q} }} & 0
\end{array}\right)
\end{eqnarray}
with $R(\mathbf{q} ) \equiv \sqrt{v_{x}\left( \mathbf{k} ^{0}\right)^{2} q_{x}^{2}+v_{y}\left( \mathbf{k} ^{0}\right)^{2} q_{y}^{2}}$ and $\theta_{  \mathbf{q} } \equiv$ $\tan ^{-1}\left[\frac{v_{y}\left( \mathbf{k}  ^{0}\right) q_{y}}{v_{x}\left( \mathbf{k}  ^{0}\right) q_{x}}\right]$. The corresponding winding number is then given by
\begin{eqnarray}
W(\bfk^0) = \frac{1}{2 \pi i} \oint d\theta_{\tilde{\bf{q}}} \partial_{\theta_{\tilde{\bf{q}}}} e ^{i \theta_{\tilde{\bf{q}}}}
=\pm 1,
\end{eqnarray}
depending on the positions of Weyl points. Note that due to the fermion-doubling theorem \cite{NIELSEN198120,RevModPhys.90.015001,PhysRevB.81.134515,doi:10.1126/sciadv.aat2374}, the sum of all topological charges needs to be zero since topological point nodes with positive and negative values come in pairs. 

In Fig. \ref{Fig:band structure and edge state}(d), we show the calculated results of topological charges for each gapless point in the momentum space. 
The bulk energy spectrum is shown in Fig. \ref{Fig:band structure and edge state}(a).
It is clear that the points of positive charge and negative charge appears alternatively in space, while the breakdown of rotational $C_4$ symmetry makes the projection of total charge to be zero on the $k_y$ axis, but non-zero on the $k_x$ axis [see Figs. \ref{Fig:band structure and edge state}(b)(c)]. It explains why a zero mode appears in the edge state when the boundary is open in the $y$ direction (hence $k_x$ is a conserved quantum number), as shown in Fig. \ref{Fig:band structure and edge state}(b).

\section{Finite-Temperature Effects}
\label{Sec: Finite Temperature}

After studying the quantum phase diagram at zero temperature, we further investigate the finite temperature effects of such a gapless topological superfluid, which has co-existing $s$- and $p$-wave order parameters. The phases at a finite temperature can be obtained by minimizing the total Helmholtz free energy, $F(T)=E(T)-TS(T)$, where within the mean-field approximation, the energy $E(T)=\langle\hat{H}_{BCS}\rangle+E_C$ and the entropy $S(T)$ has to be calculated by including the thermal distribution of the Bogoliubov quasi-particles at a temperature $T$. Here the thermal expectation value of the BCS effective Hamiltonian is given by (see Eq. (\ref{Eq:H_BCS}))
\begin{eqnarray}
 \langle \hat{H}_{BCS}\rangle 
 &=&\frac{1}{2}\sum_{\mathbf{k}}
\left[ E_{1, \mathbf{k}} \left\langle\hat{\gamma}_{1,\mathbf{k} }^{\dagger} \hat{\gamma}_{1,\mathbf{k} }\right\rangle+ E_{2, \mathbf{k}} \left\langle\hat{\gamma}_{2,\mathbf{k} }^{\dagger} \hat{\gamma}_{2,\mathbf{k} }\right\rangle
\right.
\nonumber\\
&&
\left.+E_{3, \mathbf{k}} \left\langle\hat{\gamma}_{3,\mathbf{-k}} \hat{\gamma}_{3,\mathbf{-k} }^{\dagger}\right\rangle +E_{4, \mathbf{k}} \left\langle\hat{\gamma}_{4,\mathbf{-k}} \hat{\gamma}_{4,\mathbf{-k}  }^{\dagger}\right\rangle \right] 
\nonumber
\\
&=&\frac{1}{2}\sum_{\mathbf{k}}
\left[ E_{1, \mathbf{k}} f_{1,\mathbf{k}  }+ E_{2, \mathbf{k}} f_{2,\mathbf{k}  }\right.
\nonumber\\
&&\left.+E_{3, \mathbf{k}} \left(1-f_{3,-\mathbf{k}  }\right)+E_{4, \mathbf{k}} \left(1-f_{4,-\mathbf{k}  }\right)\right],
\end{eqnarray}
where $E_{j,\mathbf{k}} (j=1, \cdots 4)$ are the four eigenvalues of the BCS Hamiltonian. The quasi-particle thermal expectation is given by $\left\langle \hat{\gamma}_{j,\mathbf{k}    }^{\dagger}  \hat{\gamma}_{j,\mathbf{k} } \right\rangle=f_{j, \mathbf{k} }$ at the finite temperature,  where $f_{j,\mathbf{k}}=\frac{1}{\mathrm{e}^{E_{j,\mathbf{k}}/ T}+1}$ is the Fermi-Dirac distribution. Similarly, the entropy could be also calculated easily from these elementary excitations:
\begin{eqnarray}
S(T)&=&- \sum_{j,\mathbf{k}}
\left[f_{j,\mathbf{k}}  \mathrm{ln} f_{j,\mathbf{k}}+\left(1-f_{j,\mathbf{k}}\right)  \mathrm{ln}\left(1-f_{j,\mathbf{k}}\right)\right]. \notag\\
\label{Eq:entropy}
\end{eqnarray}

As a result, the finite temperature order parameter (i.e.  $\Delta_0^s(T)$, $\Delta_1^s(T)$, and $\Delta_1^p(T)$) are now treated as a temperature dependent variational parameters to minimize the total Helmholtz free energy, $F(T)$, where both the constant energy, $E_C$, and the elementary excitation energy, $E_{j,\bfk}$, have such order parameter dependence as shown in Eqs. (\ref{Eq:E_c}) and (\ref{Eq:E_jk_00}) respectively.

\begin{figure}[htb]
\centering
\captionsetup{justification=raggedright,singlelinecheck=false}
\includegraphics[width=0.5\textwidth]{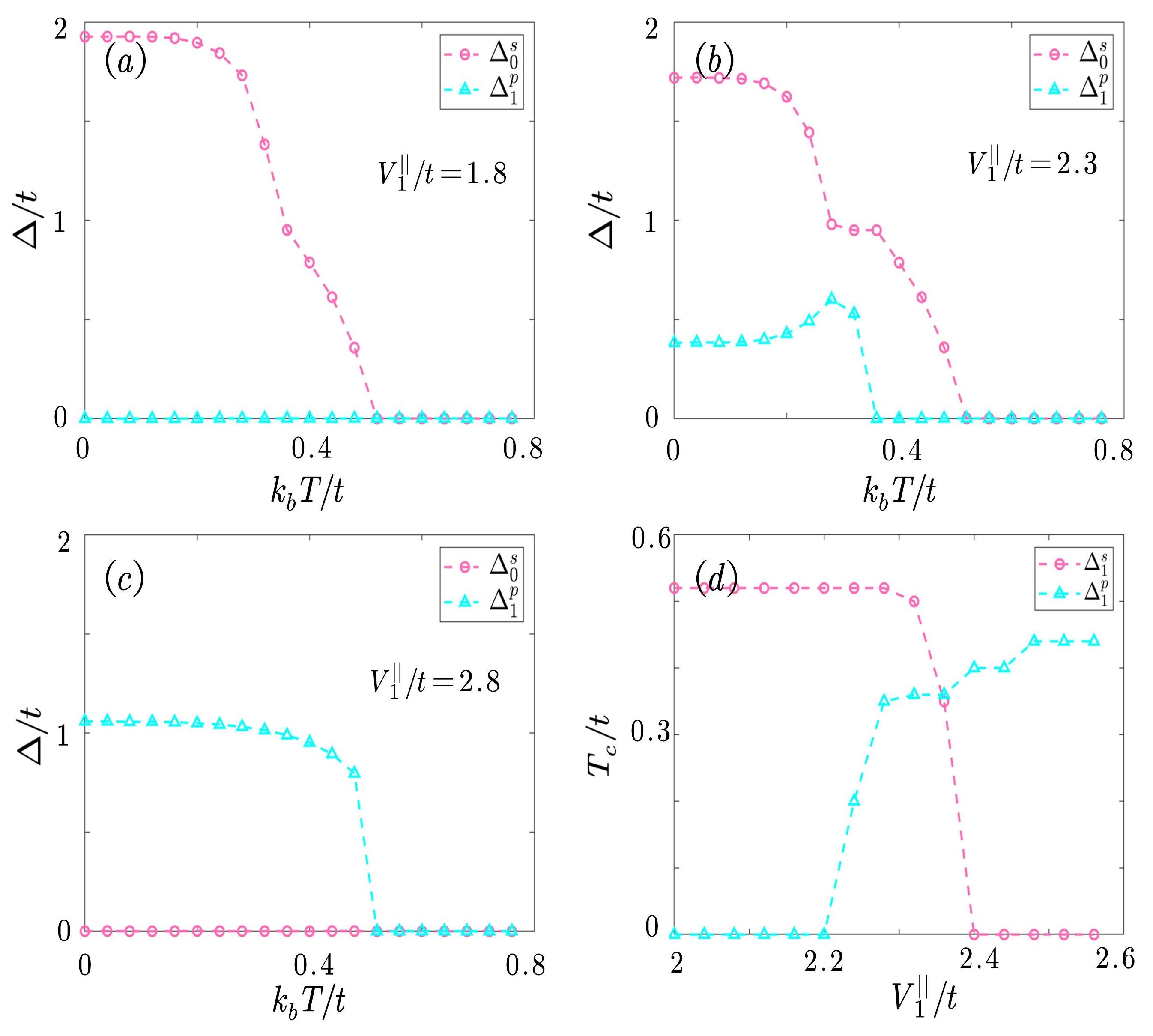}
\caption{(a) Temperature dependence of the $s-$wave pairing order parameter, $\Delta_{0}^{s}(T)$, and the $p-$wave pairing order parameter, $\Delta_{1}^{p}(T)$ in the parameter regime at the point (a) in Fig. \ref{Fig:quantum phase diagram}. (b) and (c) are the same as (a), but with different interaction strength, see Fig. \ref{Fig:quantum phase diagram}. (d) shows the critical temperature calculated as a function of $V^\|_1/t$ for $V_{0}^{\perp}/t=5$. There is a regime when both pairing exist with two critical temperatures.}
\label{Fig:Finite temperature}
\end{figure}

In Fig. \ref{Fig:Finite temperature}(a) and (c) we show the results of the order parameters, $\Delta_0^s(T)$ and $\Delta_1^p(T)$, as a function of temperature inside the gapped superfluid phase, where $\Delta_1^p(T)\to 0$ and $\Delta_1^s(T)\to 0$
in these two parameter regimes. They correspond to (a) and (c) points in Fig.~\ref{Fig:quantum phase diagram}, where the pairing phases have been selected to be $(\alpha,\phi)=(\pi,\pi)$. (Note that when only $s$- or $p$-wave phase exists, there will be no pairing phase dependence in the energy and hence no difference if using other $(\alpha,\phi)$.)
However, in the intermediate parameter regime, as shown in Fig.~\ref{Fig:Finite temperature}(b), these two order parameters could co-exist and changes in different way as a function of temperature. One could see that both $s-$ and $p-$wave order parameters change non-monotonously as the temperature increases, and then decreases to zero at different critical temperature by thermal fluctuations, quiet different from traditional superfluids with single order parameter only. 
The presence of two critical temperatures, $T_c^{s,p}$, also imply that there should be a {\it finite temperature} topological phase transitions in the intermediate temperature regime. 
In other words, the finite temperature topological phase transition indicates from a non-vanishing $p-$wave order parameter to vanishing $p-$wave order parameter in the presence 
of the coexisted $s-$wave order parameter.
Note that, $T_c^s$ could be either larger or smaller than $T_c^p$, depending on the parameters we consider within this gapless topological superfluid phase.

In Fig. \ref{Fig:Finite temperature}(d), we show how the critical temperature of the two order parameters change as a function of intra-layer nearest neighboring interaction strength, $V_{1}^{\|}$, by keeping the inter-layer on-site interaction, $V_{0}^{\perp}/t=5$ along the vertical line in Fig.~\ref{Fig:quantum phase diagram}). 
One can see that the system is dominated by $p-$wave superfluid in one side and dominated by $s-$wave superfluid in the other side. In the intermediate regime, both of them exist and the associated phase is a gapless topological superfluid as we discussed above. We emphasize that such a result is obtained by a self-consistent mean-field approximation through the minimization of the total free energy for a given relative pairing phase $(\alpha,\phi)=(\pi,\pi)$, and the co-existing result could not be obtained using other values of pairing phases, because the obtained free energy is higher than the present one. Therefore, by choosing proper pairing phases between these order parameters, the obtained gapless topological superfluid could exist not only at zero temperature but might also have highly non-trivial finite temperature effects.  

\section{Conclusion}
\label{Conclusion}

We systematically studied the ground state energy and quantum phase diagram of a 2D gapless topological superfluid. This topological system is constructed by a spin polarized fermionic atoms with a weakly coupling to Rydberg excited states in a bilayer optical lattice. Based on a self-consistent variational approximation, our numerical calculations show that the interlayer $s-$wave pairing and the intralayer $p-$wave pairing could co-exist in a parameter regime through a second order phase transition, if their relative pairing phases are modified spontaneously. 
We show that the obtained ground state is a gapless topological phases with a localized flat band in one of the edge. We further investigate the finite temperature phase diagram by minimizing the free energy. The competition between these two order parameters make a non-monotonic decrease of their order parameters, and the superfluid properties have two different critical temperatures. These interesting quantum and finite temperature many-body physics should be observable within the present experimental regimes.

\section{Acknowledgement}

We thank Chung-Yu Mou, Chi-Ting Ho and Po-Hao Chou for insightful discussions. CYH is supported by the MOST of Taiwan under Grants No. 108-2112-M-029-006-MY3. 
PYC is supported by the MOST of Taiwan under Grants No.
111-2636-M-007-008.
DWW is supported by the MOST of Taiwan under Grants No. 107-2112-M-007-019-MY3. This work is also supported by 
National Center for theoretical Sciences and by the Higher Education Sprout Project funded by the Ministry of Science and Technology and Ministry of Education in Taiwan. 

\nocite{*}
\bibliography{ref}

\providecommand{\noopsort}[1]{}\providecommand{\singleletter}[1]{#1}%
\begin{thebibliography}{49}%
\makeatletter
\providecommand \@ifxundefined [1]{%
 \@ifx{#1\undefined}
}%
\providecommand \@ifnum [1]{%
 \ifnum #1\expandafter \@firstoftwo
 \else \expandafter \@secondoftwo
 \fi
}%
\providecommand \@ifx [1]{%
 \ifx #1\expandafter \@firstoftwo
 \else \expandafter \@secondoftwo
 \fi
}%
\providecommand \natexlab [1]{#1}%
\providecommand \enquote  [1]{``#1''}%
\providecommand \bibnamefont  [1]{#1}%
\providecommand \bibfnamefont [1]{#1}%
\providecommand \citenamefont [1]{#1}%
\providecommand \href@noop [0]{\@secondoftwo}%
\providecommand \href [0]{\begingroup \@sanitize@url \@href}%
\providecommand \@href[1]{\@@startlink{#1}\@@href}%
\providecommand \@@href[1]{\endgroup#1\@@endlink}%
\providecommand \@sanitize@url [0]{\catcode `\\12\catcode `\$12\catcode
  `\&12\catcode `\#12\catcode `\^12\catcode `\_12\catcode `\%12\relax}%
\providecommand \@@startlink[1]{}%
\providecommand \@@endlink[0]{}%
\providecommand \url  [0]{\begingroup\@sanitize@url \@url }%
\providecommand \@url [1]{\endgroup\@href {#1}{\urlprefix }}%
\providecommand \urlprefix  [0]{URL }%
\providecommand \Eprint [0]{\href }%
\providecommand \doibase [0]{https://doi.org/}%
\providecommand \selectlanguage [0]{\@gobble}%
\providecommand \bibinfo  [0]{\@secondoftwo}%
\providecommand \bibfield  [0]{\@secondoftwo}%
\providecommand \translation [1]{[#1]}%
\providecommand \BibitemOpen [0]{}%
\providecommand \bibitemStop [0]{}%
\providecommand \bibitemNoStop [0]{.\EOS\space}%
\providecommand \EOS [0]{\spacefactor3000\relax}%
\providecommand \BibitemShut  [1]{\csname bibitem#1\endcsname}%
\let\auto@bib@innerbib\@empty
\bibitem [{\citenamefont {Majorana}(2008)}]{MF}%
  \BibitemOpen
  \bibfield  {author} {\bibinfo {author} {\bibfnamefont {E.}~\bibnamefont
  {Majorana}},\ }\bibfield  {title} {\bibinfo {title} {Teoria simmetrica
  dell'elettrone e del positrone},\ }\href {https://doi.org/10.1007/BF02961314}
  {\bibfield  {journal} {\bibinfo  {journal} {Il Nuovo Cimento (1924-1942)}\
  }\textbf {\bibinfo {volume} {14}},\ \bibinfo {pages} {171} (\bibinfo {year}
  {2008})}\BibitemShut {NoStop}%
\bibitem [{\citenamefont {Kitaev}(2003)}]{MFQC}%
  \BibitemOpen
  \bibfield  {author} {\bibinfo {author} {\bibfnamefont {A.}~\bibnamefont
  {Kitaev}},\ }\bibfield  {title} {\bibinfo {title} {Fault-tolerant quantum
  computation by anyons},\ }\href
  {https://doi.org/10.1016/s0003-4916(02)00018-0} {\bibfield  {journal}
  {\bibinfo  {journal} {Annals of Physics}\ }\textbf {\bibinfo {volume}
  {303}},\ \bibinfo {pages} {2–30} (\bibinfo {year} {2003})}\BibitemShut
  {NoStop}%
\bibitem [{\citenamefont {Nayak}\ \emph {et~al.}(2008)\citenamefont {Nayak},
  \citenamefont {Simon}, \citenamefont {Stern}, \citenamefont {Freedman},\ and\
  \citenamefont {Das~Sarma}}]{MFQC2}%
  \BibitemOpen
  \bibfield  {author} {\bibinfo {author} {\bibfnamefont {C.}~\bibnamefont
  {Nayak}}, \bibinfo {author} {\bibfnamefont {S.~H.}\ \bibnamefont {Simon}},
  \bibinfo {author} {\bibfnamefont {A.}~\bibnamefont {Stern}}, \bibinfo
  {author} {\bibfnamefont {M.}~\bibnamefont {Freedman}},\ and\ \bibinfo
  {author} {\bibfnamefont {S.}~\bibnamefont {Das~Sarma}},\ }\bibfield  {title}
  {\bibinfo {title} {Non-abelian anyons and topological quantum computation},\
  }\href {https://doi.org/10.1103/revmodphys.80.1083} {\bibfield  {journal}
  {\bibinfo  {journal} {Reviews of Modern Physics}\ }\textbf {\bibinfo {volume}
  {80}},\ \bibinfo {pages} {1083–1159} (\bibinfo {year} {2008})}\BibitemShut
  {NoStop}%
\bibitem [{\citenamefont {Sarma}\ \emph {et~al.}(2015)\citenamefont {Sarma},
  \citenamefont {Freedman},\ and\ \citenamefont
  {Nayak}}]{sarmaMajoranaZeroModes2015}%
  \BibitemOpen
  \bibfield  {author} {\bibinfo {author} {\bibfnamefont {S.~D.}\ \bibnamefont
  {Sarma}}, \bibinfo {author} {\bibfnamefont {M.}~\bibnamefont {Freedman}},\
  and\ \bibinfo {author} {\bibfnamefont {C.}~\bibnamefont {Nayak}},\ }\bibfield
   {title} {\bibinfo {title} {Majorana zero modes and topological quantum
  computation},\ }\href {https://doi.org/10.1038/npjqi.2015.1} {\bibfield
  {journal} {\bibinfo  {journal} {npj Quantum Information}\ }\textbf {\bibinfo
  {volume} {1}},\ \bibinfo {pages} {1} (\bibinfo {year} {2015})}\BibitemShut
  {NoStop}%
\bibitem [{\citenamefont {Qi}\ and\ \citenamefont {Zhang}(2011)}]{MFSF}%
  \BibitemOpen
  \bibfield  {author} {\bibinfo {author} {\bibfnamefont {X.-L.}\ \bibnamefont
  {Qi}}\ and\ \bibinfo {author} {\bibfnamefont {S.-C.}\ \bibnamefont {Zhang}},\
  }\bibfield  {title} {\bibinfo {title} {Topological insulators and
  superconductors},\ }\href {https://doi.org/10.1103/revmodphys.83.1057}
  {\bibfield  {journal} {\bibinfo  {journal} {Reviews of Modern Physics}\
  }\textbf {\bibinfo {volume} {83}},\ \bibinfo {pages} {1057–1110} (\bibinfo
  {year} {2011})}\BibitemShut {NoStop}%
\bibitem [{\citenamefont {Kitaev}(2001)}]{Kitaev_1D}%
  \BibitemOpen
  \bibfield  {author} {\bibinfo {author} {\bibfnamefont {A.~Y.}\ \bibnamefont
  {Kitaev}},\ }\bibfield  {title} {\bibinfo {title} {Unpaired majorana fermions
  in quantum wires},\ }\href {http://stacks.iop.org/1063-7869/44/i=10S/a=S29}
  {\bibfield  {journal} {\bibinfo  {journal} {Physics-Uspekhi}\ }\textbf
  {\bibinfo {volume} {44}},\ \bibinfo {pages} {131} (\bibinfo {year}
  {2001})}\BibitemShut {NoStop}%
\bibitem [{\citenamefont {Fu}\ and\ \citenamefont
  {Kane}(2008)}]{PhysRevLett.100.096407}%
  \BibitemOpen
  \bibfield  {author} {\bibinfo {author} {\bibfnamefont {L.}~\bibnamefont
  {Fu}}\ and\ \bibinfo {author} {\bibfnamefont {C.~L.}\ \bibnamefont {Kane}},\
  }\bibfield  {title} {\bibinfo {title} {Superconducting proximity effect and
  majorana fermions at the surface of a topological insulator},\ }\href
  {https://doi.org/10.1103/PhysRevLett.100.096407} {\bibfield  {journal}
  {\bibinfo  {journal} {Phys. Rev. Lett.}\ }\textbf {\bibinfo {volume} {100}},\
  \bibinfo {pages} {096407} (\bibinfo {year} {2008})}\BibitemShut {NoStop}%
\bibitem [{\citenamefont {Sau}\ \emph {et~al.}(2010)\citenamefont {Sau},
  \citenamefont {Lutchyn}, \citenamefont {Tewari},\ and\ \citenamefont
  {Das~Sarma}}]{S_Heterostructures}%
  \BibitemOpen
  \bibfield  {author} {\bibinfo {author} {\bibfnamefont {J.~D.}\ \bibnamefont
  {Sau}}, \bibinfo {author} {\bibfnamefont {R.~M.}\ \bibnamefont {Lutchyn}},
  \bibinfo {author} {\bibfnamefont {S.}~\bibnamefont {Tewari}},\ and\ \bibinfo
  {author} {\bibfnamefont {S.}~\bibnamefont {Das~Sarma}},\ }\bibfield  {title}
  {\bibinfo {title} {Generic new platform for topological quantum computation
  using semiconductor heterostructures},\ }\href
  {https://doi.org/10.1103/PhysRevLett.104.040502} {\bibfield  {journal}
  {\bibinfo  {journal} {Phys. Rev. Lett.}\ }\textbf {\bibinfo {volume} {104}},\
  \bibinfo {pages} {040502} (\bibinfo {year} {2010})}\BibitemShut {NoStop}%
\bibitem [{\citenamefont
  {Samuel~Reich}(2012)}]{samuelreichSolidCaseMajorana2012}%
  \BibitemOpen
  \bibfield  {author} {\bibinfo {author} {\bibfnamefont {E.}~\bibnamefont
  {Samuel~Reich}},\ }\bibfield  {title} {\bibinfo {title} {A solid case for
  {{Majorana}} fermions},\ }\href {https://doi.org/10.1038/483132a} {\bibfield
  {journal} {\bibinfo  {journal} {Nature}\ }\textbf {\bibinfo {volume} {483}},\
  \bibinfo {pages} {132} (\bibinfo {year} {2012})}\BibitemShut {NoStop}%
\bibitem [{\citenamefont {Stajic}(2020)}]{stajicLookingChiralMajoranas2020}%
  \BibitemOpen
  \bibfield  {author} {\bibinfo {author} {\bibfnamefont {J.}~\bibnamefont
  {Stajic}},\ }\bibfield  {title} {\bibinfo {title} {Looking for chiral
  majoranas},\ }\href {https://doi.org/10.1126/science.367.6473.36-m}
  {\bibfield  {journal} {\bibinfo  {journal} {Science}\ }\textbf {\bibinfo
  {volume} {367}},\ \bibinfo {pages} {36} (\bibinfo {year} {2020})}\BibitemShut
  {NoStop}%
\bibitem [{\citenamefont {Banerjee}\ \emph {et~al.}(2016)\citenamefont
  {Banerjee}, \citenamefont {Bridges}, \citenamefont {Yan}, \citenamefont
  {Aczel}, \citenamefont {Li}, \citenamefont {Stone}, \citenamefont {Granroth},
  \citenamefont {Lumsden}, \citenamefont {Yiu}, \citenamefont {Knolle},
  \citenamefont {Bhattacharjee}, \citenamefont {Kovrizhin}, \citenamefont
  {Moessner}, \citenamefont {Tennant}, \citenamefont {Mandrus},\ and\
  \citenamefont {Nagler}}]{banerjeeProximateKitaevQuantum2016a}%
  \BibitemOpen
  \bibfield  {author} {\bibinfo {author} {\bibfnamefont {A.}~\bibnamefont
  {Banerjee}}, \bibinfo {author} {\bibfnamefont {C.~A.}\ \bibnamefont
  {Bridges}}, \bibinfo {author} {\bibfnamefont {J.-Q.}\ \bibnamefont {Yan}},
  \bibinfo {author} {\bibfnamefont {A.~A.}\ \bibnamefont {Aczel}}, \bibinfo
  {author} {\bibfnamefont {L.}~\bibnamefont {Li}}, \bibinfo {author}
  {\bibfnamefont {M.~B.}\ \bibnamefont {Stone}}, \bibinfo {author}
  {\bibfnamefont {G.~E.}\ \bibnamefont {Granroth}}, \bibinfo {author}
  {\bibfnamefont {M.~D.}\ \bibnamefont {Lumsden}}, \bibinfo {author}
  {\bibfnamefont {Y.}~\bibnamefont {Yiu}}, \bibinfo {author} {\bibfnamefont
  {J.}~\bibnamefont {Knolle}}, \bibinfo {author} {\bibfnamefont
  {S.}~\bibnamefont {Bhattacharjee}}, \bibinfo {author} {\bibfnamefont {D.~L.}\
  \bibnamefont {Kovrizhin}}, \bibinfo {author} {\bibfnamefont {R.}~\bibnamefont
  {Moessner}}, \bibinfo {author} {\bibfnamefont {D.~A.}\ \bibnamefont
  {Tennant}}, \bibinfo {author} {\bibfnamefont {D.~G.}\ \bibnamefont
  {Mandrus}},\ and\ \bibinfo {author} {\bibfnamefont {S.~E.}\ \bibnamefont
  {Nagler}},\ }\bibfield  {title} {\bibinfo {title} {Proximate {{Kitaev}}
  quantum spin liquid behaviour in a honeycomb magnet},\ }\href
  {https://doi.org/10.1038/nmat4604} {\bibfield  {journal} {\bibinfo  {journal}
  {Nature Materials}\ }\textbf {\bibinfo {volume} {15}},\ \bibinfo {pages}
  {733} (\bibinfo {year} {2016})}\BibitemShut {NoStop}%
\bibitem [{\citenamefont {Wang}\ \emph {et~al.}(2018)\citenamefont {Wang},
  \citenamefont {Kong}, \citenamefont {Fan}, \citenamefont {Chen},
  \citenamefont {Zhu}, \citenamefont {Liu}, \citenamefont {Cao}, \citenamefont
  {Sun}, \citenamefont {Du}, \citenamefont {Schneeloch}, \citenamefont {Zhong},
  \citenamefont {Gu}, \citenamefont {Fu}, \citenamefont {Ding},\ and\
  \citenamefont {Gao}}]{wangEvidenceMajoranaBound2018}%
  \BibitemOpen
  \bibfield  {author} {\bibinfo {author} {\bibfnamefont {D.}~\bibnamefont
  {Wang}}, \bibinfo {author} {\bibfnamefont {L.}~\bibnamefont {Kong}}, \bibinfo
  {author} {\bibfnamefont {P.}~\bibnamefont {Fan}}, \bibinfo {author}
  {\bibfnamefont {H.}~\bibnamefont {Chen}}, \bibinfo {author} {\bibfnamefont
  {S.}~\bibnamefont {Zhu}}, \bibinfo {author} {\bibfnamefont {W.}~\bibnamefont
  {Liu}}, \bibinfo {author} {\bibfnamefont {L.}~\bibnamefont {Cao}}, \bibinfo
  {author} {\bibfnamefont {Y.}~\bibnamefont {Sun}}, \bibinfo {author}
  {\bibfnamefont {S.}~\bibnamefont {Du}}, \bibinfo {author} {\bibfnamefont
  {J.}~\bibnamefont {Schneeloch}}, \bibinfo {author} {\bibfnamefont
  {R.}~\bibnamefont {Zhong}}, \bibinfo {author} {\bibfnamefont
  {G.}~\bibnamefont {Gu}}, \bibinfo {author} {\bibfnamefont {L.}~\bibnamefont
  {Fu}}, \bibinfo {author} {\bibfnamefont {H.}~\bibnamefont {Ding}},\ and\
  \bibinfo {author} {\bibfnamefont {H.-J.}\ \bibnamefont {Gao}},\ }\bibfield
  {title} {\bibinfo {title} {Evidence for {{Majorana}} bound states in an
  iron-based superconductor},\ }\href {https://doi.org/10.1126/science.aao1797}
  {\bibfield  {journal} {\bibinfo  {journal} {Science}\ }\textbf {\bibinfo
  {volume} {362}},\ \bibinfo {pages} {333} (\bibinfo {year}
  {2018})}\BibitemShut {NoStop}%
\bibitem [{\citenamefont {Kayyalha}\ \emph {et~al.}(2020)\citenamefont
  {Kayyalha}, \citenamefont {Xiao}, \citenamefont {Zhang}, \citenamefont
  {Shin}, \citenamefont {Jiang}, \citenamefont {Wang}, \citenamefont {Zhao},
  \citenamefont {Xiao}, \citenamefont {Zhang}, \citenamefont {Fijalkowski},
  \citenamefont {Mandal}, \citenamefont {Winnerlein}, \citenamefont {Gould},
  \citenamefont {Li}, \citenamefont {Molenkamp}, \citenamefont {Chan},
  \citenamefont {Samarth},\ and\ \citenamefont
  {Chang}}]{kayyalhaAbsenceEvidenceChiral2020}%
  \BibitemOpen
  \bibfield  {author} {\bibinfo {author} {\bibfnamefont {M.}~\bibnamefont
  {Kayyalha}}, \bibinfo {author} {\bibfnamefont {D.}~\bibnamefont {Xiao}},
  \bibinfo {author} {\bibfnamefont {R.}~\bibnamefont {Zhang}}, \bibinfo
  {author} {\bibfnamefont {J.}~\bibnamefont {Shin}}, \bibinfo {author}
  {\bibfnamefont {J.}~\bibnamefont {Jiang}}, \bibinfo {author} {\bibfnamefont
  {F.}~\bibnamefont {Wang}}, \bibinfo {author} {\bibfnamefont {Y.-F.}\
  \bibnamefont {Zhao}}, \bibinfo {author} {\bibfnamefont {R.}~\bibnamefont
  {Xiao}}, \bibinfo {author} {\bibfnamefont {L.}~\bibnamefont {Zhang}},
  \bibinfo {author} {\bibfnamefont {K.~M.}\ \bibnamefont {Fijalkowski}},
  \bibinfo {author} {\bibfnamefont {P.}~\bibnamefont {Mandal}}, \bibinfo
  {author} {\bibfnamefont {M.}~\bibnamefont {Winnerlein}}, \bibinfo {author}
  {\bibfnamefont {C.}~\bibnamefont {Gould}}, \bibinfo {author} {\bibfnamefont
  {Q.}~\bibnamefont {Li}}, \bibinfo {author} {\bibfnamefont {L.~W.}\
  \bibnamefont {Molenkamp}}, \bibinfo {author} {\bibfnamefont {M.~H.~W.}\
  \bibnamefont {Chan}}, \bibinfo {author} {\bibfnamefont {N.}~\bibnamefont
  {Samarth}},\ and\ \bibinfo {author} {\bibfnamefont {C.-Z.}\ \bibnamefont
  {Chang}},\ }\bibfield  {title} {\bibinfo {title} {Absence of evidence for
  chiral {{Majorana}} modes in quantum anomalous {{Hall}}-superconductor
  devices},\ }\href {https://doi.org/10.1126/science.aax6361} {\bibfield
  {journal} {\bibinfo  {journal} {Science}\ }\textbf {\bibinfo {volume}
  {367}},\ \bibinfo {pages} {64} (\bibinfo {year} {2020})}\BibitemShut
  {NoStop}%
\bibitem [{\citenamefont
  {Castelvecchi}(2021)}]{castelvecchiEvidenceElusiveMajorana2021}%
  \BibitemOpen
  \bibfield  {author} {\bibinfo {author} {\bibfnamefont {D.}~\bibnamefont
  {Castelvecchi}},\ }\bibfield  {title} {\bibinfo {title} {Evidence of elusive
  majorana particle dies - but computing hope lives on},\ }\href
  {https://doi.org/10.1038/d41586-021-00612-z} {\bibfield  {journal} {\bibinfo
  {journal} {Nature}\ }\textbf {\bibinfo {volume} {591}},\ \bibinfo {pages}
  {354} (\bibinfo {year} {2021})}\BibitemShut {NoStop}%
\bibitem [{\citenamefont {Jiang}\ \emph {et~al.}(2011)\citenamefont {Jiang},
  \citenamefont {Kitagawa}, \citenamefont {Alicea}, \citenamefont {Akhmerov},
  \citenamefont {Pekker}, \citenamefont {Refael}, \citenamefont {Cirac},
  \citenamefont {Demler}, \citenamefont {Lukin},\ and\ \citenamefont
  {Zoller}}]{MF_Cold}%
  \BibitemOpen
  \bibfield  {author} {\bibinfo {author} {\bibfnamefont {L.}~\bibnamefont
  {Jiang}}, \bibinfo {author} {\bibfnamefont {T.}~\bibnamefont {Kitagawa}},
  \bibinfo {author} {\bibfnamefont {J.}~\bibnamefont {Alicea}}, \bibinfo
  {author} {\bibfnamefont {A.~R.}\ \bibnamefont {Akhmerov}}, \bibinfo {author}
  {\bibfnamefont {D.}~\bibnamefont {Pekker}}, \bibinfo {author} {\bibfnamefont
  {G.}~\bibnamefont {Refael}}, \bibinfo {author} {\bibfnamefont {J.~I.}\
  \bibnamefont {Cirac}}, \bibinfo {author} {\bibfnamefont {E.}~\bibnamefont
  {Demler}}, \bibinfo {author} {\bibfnamefont {M.~D.}\ \bibnamefont {Lukin}},\
  and\ \bibinfo {author} {\bibfnamefont {P.}~\bibnamefont {Zoller}},\
  }\bibfield  {title} {\bibinfo {title} {Majorana fermions in equilibrium and
  in driven cold-atom quantum wires},\ }\bibfield  {journal} {\bibinfo
  {journal} {Physical Review Letters}\ }\textbf {\bibinfo {volume} {106}},\
  \href {https://doi.org/10.1103/physrevlett.106.220402}
  {10.1103/physrevlett.106.220402} (\bibinfo {year} {2011})\BibitemShut
  {NoStop}%
\bibitem [{\citenamefont {Tewari}\ \emph {et~al.}(2007)\citenamefont {Tewari},
  \citenamefont {Das~Sarma}, \citenamefont {Nayak}, \citenamefont {Zhang},\
  and\ \citenamefont {Zoller}}]{MZM_Fermionic_Cold_Atoms}%
  \BibitemOpen
  \bibfield  {author} {\bibinfo {author} {\bibfnamefont {S.}~\bibnamefont
  {Tewari}}, \bibinfo {author} {\bibfnamefont {S.}~\bibnamefont {Das~Sarma}},
  \bibinfo {author} {\bibfnamefont {C.}~\bibnamefont {Nayak}}, \bibinfo
  {author} {\bibfnamefont {C.}~\bibnamefont {Zhang}},\ and\ \bibinfo {author}
  {\bibfnamefont {P.}~\bibnamefont {Zoller}},\ }\bibfield  {title} {\bibinfo
  {title} {Quantum computation using vortices and majorana zero modes of a
  ${p}_{x}+i{p}_{y}$ superfluid of fermionic cold atoms},\ }\href
  {https://doi.org/10.1103/PhysRevLett.98.010506} {\bibfield  {journal}
  {\bibinfo  {journal} {Phys. Rev. Lett.}\ }\textbf {\bibinfo {volume} {98}},\
  \bibinfo {pages} {010506} (\bibinfo {year} {2007})}\BibitemShut {NoStop}%
\bibitem [{\citenamefont {Shermadini}\ \emph {et~al.}(2011)\citenamefont
  {Shermadini}, \citenamefont {Krzton-Maziopa}, \citenamefont {Bendele},
  \citenamefont {Khasanov}, \citenamefont {Luetkens}, \citenamefont {Conder},
  \citenamefont {Pomjakushina}, \citenamefont {Weyeneth}, \citenamefont
  {Pomjakushin}, \citenamefont {Bossen},\ and\ \citenamefont
  {Amato}}]{MF_Cold_Atom_Quantum_Wires}%
  \BibitemOpen
  \bibfield  {author} {\bibinfo {author} {\bibfnamefont {Z.}~\bibnamefont
  {Shermadini}}, \bibinfo {author} {\bibfnamefont {A.}~\bibnamefont
  {Krzton-Maziopa}}, \bibinfo {author} {\bibfnamefont {M.}~\bibnamefont
  {Bendele}}, \bibinfo {author} {\bibfnamefont {R.}~\bibnamefont {Khasanov}},
  \bibinfo {author} {\bibfnamefont {H.}~\bibnamefont {Luetkens}}, \bibinfo
  {author} {\bibfnamefont {K.}~\bibnamefont {Conder}}, \bibinfo {author}
  {\bibfnamefont {E.}~\bibnamefont {Pomjakushina}}, \bibinfo {author}
  {\bibfnamefont {S.}~\bibnamefont {Weyeneth}}, \bibinfo {author}
  {\bibfnamefont {V.}~\bibnamefont {Pomjakushin}}, \bibinfo {author}
  {\bibfnamefont {O.}~\bibnamefont {Bossen}},\ and\ \bibinfo {author}
  {\bibfnamefont {A.}~\bibnamefont {Amato}},\ }\bibfield  {title} {\bibinfo
  {title} {Coexistence of magnetism and superconductivity in the iron-based
  compound ${\mathrm{cs}}_{0.8}({\mathrm{fese}}_{0.98}{)}_{2}$},\ }\href
  {https://doi.org/10.1103/PhysRevLett.106.117602} {\bibfield  {journal}
  {\bibinfo  {journal} {Phys. Rev. Lett.}\ }\textbf {\bibinfo {volume} {106}},\
  \bibinfo {pages} {117602} (\bibinfo {year} {2011})}\BibitemShut {NoStop}%
\bibitem [{\citenamefont {Liu}\ \emph {et~al.}(2014)\citenamefont {Liu},
  \citenamefont {Law},\ and\ \citenamefont {Ng}}]{2D_Spin_Orbit}%
  \BibitemOpen
  \bibfield  {author} {\bibinfo {author} {\bibfnamefont {X.-J.}\ \bibnamefont
  {Liu}}, \bibinfo {author} {\bibfnamefont {K.~T.}\ \bibnamefont {Law}},\ and\
  \bibinfo {author} {\bibfnamefont {T.~K.}\ \bibnamefont {Ng}},\ }\bibfield
  {title} {\bibinfo {title} {Realization of 2d spin-orbit interaction and
  exotic topological orders in cold atoms},\ }\href
  {https://doi.org/10.1103/PhysRevLett.112.086401} {\bibfield  {journal}
  {\bibinfo  {journal} {Phys. Rev. Lett.}\ }\textbf {\bibinfo {volume} {112}},\
  \bibinfo {pages} {086401} (\bibinfo {year} {2014})}\BibitemShut {NoStop}%
\bibitem [{\citenamefont {Sato}\ \emph {et~al.}(2009)\citenamefont {Sato},
  \citenamefont {Takahashi},\ and\ \citenamefont {Fujimoto}}]{s_Wave_SF}%
  \BibitemOpen
  \bibfield  {author} {\bibinfo {author} {\bibfnamefont {M.}~\bibnamefont
  {Sato}}, \bibinfo {author} {\bibfnamefont {Y.}~\bibnamefont {Takahashi}},\
  and\ \bibinfo {author} {\bibfnamefont {S.}~\bibnamefont {Fujimoto}},\
  }\bibfield  {title} {\bibinfo {title} {Non-abelian topological order in
  $s$-wave superfluids of ultracold fermionic atoms},\ }\href
  {https://doi.org/10.1103/PhysRevLett.103.020401} {\bibfield  {journal}
  {\bibinfo  {journal} {Phys. Rev. Lett.}\ }\textbf {\bibinfo {volume} {103}},\
  \bibinfo {pages} {020401} (\bibinfo {year} {2009})}\BibitemShut {NoStop}%
\bibitem [{\citenamefont {Huang}\ \emph {et~al.}(2021)\citenamefont {Huang},
  \citenamefont {Zhuang}, \citenamefont {Chang},\ and\ \citenamefont
  {Wang}}]{huang2021twodimensional}%
  \BibitemOpen
  \bibfield  {author} {\bibinfo {author} {\bibfnamefont {C.-Y.}\ \bibnamefont
  {Huang}}, \bibinfo {author} {\bibfnamefont {J.}~\bibnamefont {Zhuang}},
  \bibinfo {author} {\bibfnamefont {P.-Y.}\ \bibnamefont {Chang}},\ and\
  \bibinfo {author} {\bibfnamefont {D.-W.}\ \bibnamefont {Wang}},\ }\href@noop
  {} {\bibinfo {title} {Two-dimensional paired topological superfluids of
  rydberg fermi gases}} (\bibinfo {year} {2021}),\ \Eprint
  {https://arxiv.org/abs/2112.14027} {arXiv:2112.14027 [cond-mat.supr-con]}
  \BibitemShut {NoStop}%
\bibitem [{\citenamefont {Pikovski}\ \emph {et~al.}(2010)\citenamefont
  {Pikovski}, \citenamefont {Klawunn}, \citenamefont {Shlyapnikov},\ and\
  \citenamefont {Santos}}]{Pikovski_2010}%
  \BibitemOpen
  \bibfield  {author} {\bibinfo {author} {\bibfnamefont {A.}~\bibnamefont
  {Pikovski}}, \bibinfo {author} {\bibfnamefont {M.}~\bibnamefont {Klawunn}},
  \bibinfo {author} {\bibfnamefont {G.~V.}\ \bibnamefont {Shlyapnikov}},\ and\
  \bibinfo {author} {\bibfnamefont {L.}~\bibnamefont {Santos}},\ }\bibfield
  {title} {\bibinfo {title} {Interlayer superfluidity in bilayer systems of
  fermionic polar molecules},\ }\bibfield  {journal} {\bibinfo  {journal}
  {Physical Review Letters}\ }\textbf {\bibinfo {volume} {105}},\ \href
  {https://doi.org/10.1103/physrevlett.105.215302}
  {10.1103/physrevlett.105.215302} (\bibinfo {year} {2010})\BibitemShut
  {NoStop}%
\bibitem [{\citenamefont {Zinner}\ \emph {et~al.}(2012)\citenamefont {Zinner},
  \citenamefont {Wunsch}, \citenamefont {Pekker},\ and\ \citenamefont
  {Wang}}]{Zinner_2012}%
  \BibitemOpen
  \bibfield  {author} {\bibinfo {author} {\bibfnamefont {N.~T.}\ \bibnamefont
  {Zinner}}, \bibinfo {author} {\bibfnamefont {B.}~\bibnamefont {Wunsch}},
  \bibinfo {author} {\bibfnamefont {D.}~\bibnamefont {Pekker}},\ and\ \bibinfo
  {author} {\bibfnamefont {D.-W.}\ \bibnamefont {Wang}},\ }\bibfield  {title}
  {\bibinfo {title} {{BCS}-{BEC} crossover in bilayers of cold fermionic polar
  molecules},\ }\bibfield  {journal} {\bibinfo  {journal} {Physical Review A}\
  }\textbf {\bibinfo {volume} {85}},\ \href
  {https://doi.org/10.1103/physreva.85.013603} {10.1103/physreva.85.013603}
  (\bibinfo {year} {2012})\BibitemShut {NoStop}%
\bibitem [{\citenamefont {Babadi}\ and\ \citenamefont
  {Demler}(2011)}]{Babadi_2011}%
  \BibitemOpen
  \bibfield  {author} {\bibinfo {author} {\bibfnamefont {M.}~\bibnamefont
  {Babadi}}\ and\ \bibinfo {author} {\bibfnamefont {E.}~\bibnamefont
  {Demler}},\ }\bibfield  {title} {\bibinfo {title} {Collective phenomena in a
  quasi-two-dimensional system of fermionic polar molecules: Band
  renormalization and excitons},\ }\bibfield  {journal} {\bibinfo  {journal}
  {Physical Review A}\ }\textbf {\bibinfo {volume} {84}},\ \href
  {https://doi.org/10.1103/physreva.84.033636} {10.1103/physreva.84.033636}
  (\bibinfo {year} {2011})\BibitemShut {NoStop}%
\bibitem [{\citenamefont {Baranov}\ \emph {et~al.}(2011)\citenamefont
  {Baranov}, \citenamefont {Micheli}, \citenamefont {Ronen},\ and\
  \citenamefont {Zoller}}]{Baranov_2011}%
  \BibitemOpen
  \bibfield  {author} {\bibinfo {author} {\bibfnamefont {M.~A.}\ \bibnamefont
  {Baranov}}, \bibinfo {author} {\bibfnamefont {A.}~\bibnamefont {Micheli}},
  \bibinfo {author} {\bibfnamefont {S.}~\bibnamefont {Ronen}},\ and\ \bibinfo
  {author} {\bibfnamefont {P.}~\bibnamefont {Zoller}},\ }\bibfield  {title}
  {\bibinfo {title} {Bilayer superfluidity of fermionic polar molecules:
  Many-body effects},\ }\bibfield  {journal} {\bibinfo  {journal} {Physical
  Review A}\ }\textbf {\bibinfo {volume} {83}},\ \href
  {https://doi.org/10.1103/physreva.83.043602} {10.1103/physreva.83.043602}
  (\bibinfo {year} {2011})\BibitemShut {NoStop}%
\bibitem [{\citenamefont {Cinti}\ \emph {et~al.}(2017)\citenamefont {Cinti},
  \citenamefont {Wang},\ and\ \citenamefont {Boninsegni}}]{Cinti_2017}%
  \BibitemOpen
  \bibfield  {author} {\bibinfo {author} {\bibfnamefont {F.}~\bibnamefont
  {Cinti}}, \bibinfo {author} {\bibfnamefont {D.-W.}\ \bibnamefont {Wang}},\
  and\ \bibinfo {author} {\bibfnamefont {M.}~\bibnamefont {Boninsegni}},\
  }\bibfield  {title} {\bibinfo {title} {Phases of dipolar bosons in a bilayer
  geometry},\ }\bibfield  {journal} {\bibinfo  {journal} {Physical Review A}\
  }\textbf {\bibinfo {volume} {95}},\ \href
  {https://doi.org/10.1103/physreva.95.023622} {10.1103/physreva.95.023622}
  (\bibinfo {year} {2017})\BibitemShut {NoStop}%
\bibitem [{\citenamefont {Potter}\ \emph {et~al.}(2010)\citenamefont {Potter},
  \citenamefont {Berg}, \citenamefont {Wang}, \citenamefont {Halperin},\ and\
  \citenamefont {Demler}}]{Potter_2010}%
  \BibitemOpen
  \bibfield  {author} {\bibinfo {author} {\bibfnamefont {A.~C.}\ \bibnamefont
  {Potter}}, \bibinfo {author} {\bibfnamefont {E.}~\bibnamefont {Berg}},
  \bibinfo {author} {\bibfnamefont {D.-W.}\ \bibnamefont {Wang}}, \bibinfo
  {author} {\bibfnamefont {B.~I.}\ \bibnamefont {Halperin}},\ and\ \bibinfo
  {author} {\bibfnamefont {E.}~\bibnamefont {Demler}},\ }\bibfield  {title}
  {\bibinfo {title} {Superfluidity and dimerization in a multilayered system of
  fermionic polar molecules},\ }\bibfield  {journal} {\bibinfo  {journal}
  {Physical Review Letters}\ }\textbf {\bibinfo {volume} {105}},\ \href
  {https://doi.org/10.1103/physrevlett.105.220406}
  {10.1103/physrevlett.105.220406} (\bibinfo {year} {2010})\BibitemShut
  {NoStop}%
\bibitem [{\citenamefont {Huang}\ \emph {et~al.}(2019)\citenamefont {Huang},
  \citenamefont {Lin}, \citenamefont {Lee},\ and\ \citenamefont
  {Wang}}]{PhysRevA.99.043624}%
  \BibitemOpen
  \bibfield  {author} {\bibinfo {author} {\bibfnamefont {C.-Y.}\ \bibnamefont
  {Huang}}, \bibinfo {author} {\bibfnamefont {Y.-T.}\ \bibnamefont {Lin}},
  \bibinfo {author} {\bibfnamefont {H.}~\bibnamefont {Lee}},\ and\ \bibinfo
  {author} {\bibfnamefont {D.-W.}\ \bibnamefont {Wang}},\ }\bibfield  {title}
  {\bibinfo {title} {Quantum degenerate majorana surface zero modes in
  two-dimensional space},\ }\href {https://doi.org/10.1103/PhysRevA.99.043624}
  {\bibfield  {journal} {\bibinfo  {journal} {Phys. Rev. A}\ }\textbf {\bibinfo
  {volume} {99}},\ \bibinfo {pages} {043624} (\bibinfo {year}
  {2019})}\BibitemShut {NoStop}%
\bibitem [{\citenamefont {Henkel}\ \emph {et~al.}(2010)\citenamefont {Henkel},
  \citenamefont {Nath},\ and\ \citenamefont
  {Pohl}}]{henkelThreeDimensionalRotonExcitations2010}%
  \BibitemOpen
  \bibfield  {author} {\bibinfo {author} {\bibfnamefont {N.}~\bibnamefont
  {Henkel}}, \bibinfo {author} {\bibfnamefont {R.}~\bibnamefont {Nath}},\ and\
  \bibinfo {author} {\bibfnamefont {T.}~\bibnamefont {Pohl}},\ }\bibfield
  {title} {\bibinfo {title} {Three-{{Dimensional Roton Excitations}} and
  {{Supersolid Formation}} in {{Rydberg}}-{{Excited Bose}}-{{Einstein
  Condensates}}},\ }\href {https://doi.org/10.1103/PhysRevLett.104.195302}
  {\bibfield  {journal} {\bibinfo  {journal} {Physical Review Letters}\
  }\textbf {\bibinfo {volume} {104}},\ \bibinfo {pages} {195302} (\bibinfo
  {year} {2010})}\BibitemShut {NoStop}%
\bibitem [{\citenamefont {Honer}\ \emph {et~al.}(2010)\citenamefont {Honer},
  \citenamefont {Weimer}, \citenamefont {Pfau},\ and\ \citenamefont
  {B{\"u}chler}}]{honerCollectiveManyBodyInteraction2010}%
  \BibitemOpen
  \bibfield  {author} {\bibinfo {author} {\bibfnamefont {J.}~\bibnamefont
  {Honer}}, \bibinfo {author} {\bibfnamefont {H.}~\bibnamefont {Weimer}},
  \bibinfo {author} {\bibfnamefont {T.}~\bibnamefont {Pfau}},\ and\ \bibinfo
  {author} {\bibfnamefont {H.~P.}\ \bibnamefont {B{\"u}chler}},\ }\bibfield
  {title} {\bibinfo {title} {Collective {{Many}}-{{Body Interaction}} in
  {{Rydberg Dressed Atoms}}},\ }\href
  {https://doi.org/10.1103/PhysRevLett.105.160404} {\bibfield  {journal}
  {\bibinfo  {journal} {Physical Review Letters}\ }\textbf {\bibinfo {volume}
  {105}},\ \bibinfo {pages} {160404} (\bibinfo {year} {2010})}\BibitemShut
  {NoStop}%
\bibitem [{\citenamefont {Li}\ \emph {et~al.}(2012)\citenamefont {Li},
  \citenamefont {Hamadeh},\ and\ \citenamefont
  {Lesanovsky}}]{liProbingInteractionRydbergdressed2012}%
  \BibitemOpen
  \bibfield  {author} {\bibinfo {author} {\bibfnamefont {W.}~\bibnamefont
  {Li}}, \bibinfo {author} {\bibfnamefont {L.}~\bibnamefont {Hamadeh}},\ and\
  \bibinfo {author} {\bibfnamefont {I.}~\bibnamefont {Lesanovsky}},\ }\bibfield
   {title} {\bibinfo {title} {Probing the interaction between
  {{Rydberg}}-dressed atoms through interference},\ }\href
  {https://doi.org/10.1103/PhysRevA.85.053615} {\bibfield  {journal} {\bibinfo
  {journal} {Physical Review A}\ }\textbf {\bibinfo {volume} {85}},\ \bibinfo
  {pages} {053615} (\bibinfo {year} {2012})}\BibitemShut {NoStop}%
\bibitem [{\citenamefont {P{\l}odzie{\'n}}\ \emph {et~al.}(2017)\citenamefont
  {P{\l}odzie{\'n}}, \citenamefont {Lochead}, \citenamefont {{de Hond}},
  \citenamefont {{van Druten}},\ and\ \citenamefont
  {Kokkelmans}}]{plodzienRydbergDressingOnedimensional2017}%
  \BibitemOpen
  \bibfield  {author} {\bibinfo {author} {\bibfnamefont {M.}~\bibnamefont
  {P{\l}odzie{\'n}}}, \bibinfo {author} {\bibfnamefont {G.}~\bibnamefont
  {Lochead}}, \bibinfo {author} {\bibfnamefont {J.}~\bibnamefont {{de Hond}}},
  \bibinfo {author} {\bibfnamefont {N.~J.}\ \bibnamefont {{van Druten}}},\ and\
  \bibinfo {author} {\bibfnamefont {S.}~\bibnamefont {Kokkelmans}},\ }\bibfield
   {title} {\bibinfo {title} {Rydberg dressing of a one-dimensional
  {{Bose}}-{{Einstein}} condensate},\ }\href
  {https://doi.org/10.1103/PhysRevA.95.043606} {\bibfield  {journal} {\bibinfo
  {journal} {Physical Review A}\ }\textbf {\bibinfo {volume} {95}},\ \bibinfo
  {pages} {043606} (\bibinfo {year} {2017})}\BibitemShut {NoStop}%
\bibitem [{\citenamefont {Pupillo}\ \emph {et~al.}(2010)\citenamefont
  {Pupillo}, \citenamefont {Micheli}, \citenamefont {Boninsegni}, \citenamefont
  {Lesanovsky},\ and\ \citenamefont
  {Zoller}}]{pupilloStronglyCorrelatedGases2010}%
  \BibitemOpen
  \bibfield  {author} {\bibinfo {author} {\bibfnamefont {G.}~\bibnamefont
  {Pupillo}}, \bibinfo {author} {\bibfnamefont {A.}~\bibnamefont {Micheli}},
  \bibinfo {author} {\bibfnamefont {M.}~\bibnamefont {Boninsegni}}, \bibinfo
  {author} {\bibfnamefont {I.}~\bibnamefont {Lesanovsky}},\ and\ \bibinfo
  {author} {\bibfnamefont {P.}~\bibnamefont {Zoller}},\ }\bibfield  {title}
  {\bibinfo {title} {Strongly {{Correlated Gases}} of {{Rydberg}}-{{Dressed
  Atoms}}: {{Quantum}} and {{Classical Dynamics}}},\ }\href
  {https://doi.org/10.1103/PhysRevLett.104.223002} {\bibfield  {journal}
  {\bibinfo  {journal} {Physical Review Letters}\ }\textbf {\bibinfo {volume}
  {104}},\ \bibinfo {pages} {223002} (\bibinfo {year} {2010})}\BibitemShut
  {NoStop}%
\bibitem [{\citenamefont {Tong}\ \emph {et~al.}(2004)\citenamefont {Tong},
  \citenamefont {Farooqi}, \citenamefont {Stanojevic}, \citenamefont
  {Krishnan}, \citenamefont {Zhang}, \citenamefont {C{\^o}t{\'e}},
  \citenamefont {Eyler},\ and\ \citenamefont
  {Gould}}]{tongLocalBlockadeRydberg2004}%
  \BibitemOpen
  \bibfield  {author} {\bibinfo {author} {\bibfnamefont {D.}~\bibnamefont
  {Tong}}, \bibinfo {author} {\bibfnamefont {S.~M.}\ \bibnamefont {Farooqi}},
  \bibinfo {author} {\bibfnamefont {J.}~\bibnamefont {Stanojevic}}, \bibinfo
  {author} {\bibfnamefont {S.}~\bibnamefont {Krishnan}}, \bibinfo {author}
  {\bibfnamefont {Y.~P.}\ \bibnamefont {Zhang}}, \bibinfo {author}
  {\bibfnamefont {R.}~\bibnamefont {C{\^o}t{\'e}}}, \bibinfo {author}
  {\bibfnamefont {E.~E.}\ \bibnamefont {Eyler}},\ and\ \bibinfo {author}
  {\bibfnamefont {P.~L.}\ \bibnamefont {Gould}},\ }\bibfield  {title} {\bibinfo
  {title} {Local {{Blockade}} of {{Rydberg Excitation}} in an {{Ultracold
  Gas}}},\ }\href {https://doi.org/10.1103/PhysRevLett.93.063001} {\bibfield
  {journal} {\bibinfo  {journal} {Physical Review Letters}\ }\textbf {\bibinfo
  {volume} {93}},\ \bibinfo {pages} {063001} (\bibinfo {year}
  {2004})}\BibitemShut {NoStop}%
\bibitem [{\citenamefont {Saffman}\ \emph {et~al.}(2010)\citenamefont
  {Saffman}, \citenamefont {Walker},\ and\ \citenamefont
  {M{\o}lmer}}]{saffmanQuantumInformationRydberg2010}%
  \BibitemOpen
  \bibfield  {author} {\bibinfo {author} {\bibfnamefont {M.}~\bibnamefont
  {Saffman}}, \bibinfo {author} {\bibfnamefont {T.~G.}\ \bibnamefont
  {Walker}},\ and\ \bibinfo {author} {\bibfnamefont {K.}~\bibnamefont
  {M{\o}lmer}},\ }\bibfield  {title} {\bibinfo {title} {Quantum information
  with {{Rydberg}} atoms},\ }\href {https://doi.org/10.1103/RevModPhys.82.2313}
  {\bibfield  {journal} {\bibinfo  {journal} {Reviews of Modern Physics}\
  }\textbf {\bibinfo {volume} {82}},\ \bibinfo {pages} {2313} (\bibinfo {year}
  {2010})}\BibitemShut {NoStop}%
\bibitem [{\citenamefont {Browaeys}\ \emph {et~al.}(2016)\citenamefont
  {Browaeys}, \citenamefont {Barredo},\ and\ \citenamefont
  {Lahaye}}]{browaeysExperimentalInvestigationsDipole2016}%
  \BibitemOpen
  \bibfield  {author} {\bibinfo {author} {\bibfnamefont {A.}~\bibnamefont
  {Browaeys}}, \bibinfo {author} {\bibfnamefont {D.}~\bibnamefont {Barredo}},\
  and\ \bibinfo {author} {\bibfnamefont {T.}~\bibnamefont {Lahaye}},\
  }\bibfield  {title} {\bibinfo {title} {Experimental investigations of
  dipole\textendash dipole interactions between a few {{Rydberg}} atoms},\
  }\href {https://doi.org/10.1088/0953-4075/49/15/152001} {\bibfield  {journal}
  {\bibinfo  {journal} {Journal of Physics B: Atomic, Molecular and Optical
  Physics}\ }\textbf {\bibinfo {volume} {49}},\ \bibinfo {pages} {152001}
  (\bibinfo {year} {2016})}\BibitemShut {NoStop}%
\bibitem [{\citenamefont {Lee}\ \emph {et~al.}(2017)\citenamefont {Lee},
  \citenamefont {Matveenko}, \citenamefont {Wang},\ and\ \citenamefont
  {Shlyapnikov}}]{PhysRevA.96.061602}%
  \BibitemOpen
  \bibfield  {author} {\bibinfo {author} {\bibfnamefont {H.}~\bibnamefont
  {Lee}}, \bibinfo {author} {\bibfnamefont {S.~I.}\ \bibnamefont {Matveenko}},
  \bibinfo {author} {\bibfnamefont {D.-W.}\ \bibnamefont {Wang}},\ and\
  \bibinfo {author} {\bibfnamefont {G.~V.}\ \bibnamefont {Shlyapnikov}},\
  }\bibfield  {title} {\bibinfo {title} {Fulde-ferrell-larkin-ovchinnikov state
  in bilayer dipolar systems},\ }\href
  {https://doi.org/10.1103/PhysRevA.96.061602} {\bibfield  {journal} {\bibinfo
  {journal} {Phys. Rev. A}\ }\textbf {\bibinfo {volume} {96}},\ \bibinfo
  {pages} {061602} (\bibinfo {year} {2017})}\BibitemShut {NoStop}%
\bibitem [{\citenamefont {Anderson}(1958)}]{Anderson1958}%
  \BibitemOpen
  \bibfield  {author} {\bibinfo {author} {\bibfnamefont {P.~W.}\ \bibnamefont
  {Anderson}},\ }\bibfield  {title} {\bibinfo {title} {Random-phase
  approximation in the theory of superconductivity},\ }\href
  {https://doi.org/10.1103/PhysRev.112.1900} {\bibfield  {journal} {\bibinfo
  {journal} {Phys. Rev.}\ }\textbf {\bibinfo {volume} {112}},\ \bibinfo {pages}
  {1900} (\bibinfo {year} {1958})}\BibitemShut {NoStop}%
\bibitem [{\citenamefont {Matsuura}\ \emph {et~al.}(2013)\citenamefont
  {Matsuura}, \citenamefont {Chang}, \citenamefont {Schnyder},\ and\
  \citenamefont {Ryu}}]{Matsuura_2013}%
  \BibitemOpen
  \bibfield  {author} {\bibinfo {author} {\bibfnamefont {S.}~\bibnamefont
  {Matsuura}}, \bibinfo {author} {\bibfnamefont {P.-Y.}\ \bibnamefont {Chang}},
  \bibinfo {author} {\bibfnamefont {A.~P.}\ \bibnamefont {Schnyder}},\ and\
  \bibinfo {author} {\bibfnamefont {S.}~\bibnamefont {Ryu}},\ }\bibfield
  {title} {\bibinfo {title} {Protected boundary states in gapless topological
  phases},\ }\href {https://doi.org/10.1088/1367-2630/15/6/065001} {\bibfield
  {journal} {\bibinfo  {journal} {New Journal of Physics}\ }\textbf {\bibinfo
  {volume} {15}},\ \bibinfo {pages} {065001} (\bibinfo {year}
  {2013})}\BibitemShut {NoStop}%
\bibitem [{\citenamefont {Sato}(2006)}]{PhysRevB.73.214502}%
  \BibitemOpen
  \bibfield  {author} {\bibinfo {author} {\bibfnamefont {M.}~\bibnamefont
  {Sato}},\ }\bibfield  {title} {\bibinfo {title} {Nodal structure of
  superconductors with time-reversal invariance and ${\mathbf{z}}_{2}$
  topological number},\ }\href {https://doi.org/10.1103/PhysRevB.73.214502}
  {\bibfield  {journal} {\bibinfo  {journal} {Phys. Rev. B}\ }\textbf {\bibinfo
  {volume} {73}},\ \bibinfo {pages} {214502} (\bibinfo {year}
  {2006})}\BibitemShut {NoStop}%
\bibitem [{\citenamefont {Sato}\ \emph {et~al.}(2011)\citenamefont {Sato},
  \citenamefont {Tanaka}, \citenamefont {Yada},\ and\ \citenamefont
  {Yokoyama}}]{PhysRevB.83.224511}%
  \BibitemOpen
  \bibfield  {author} {\bibinfo {author} {\bibfnamefont {M.}~\bibnamefont
  {Sato}}, \bibinfo {author} {\bibfnamefont {Y.}~\bibnamefont {Tanaka}},
  \bibinfo {author} {\bibfnamefont {K.}~\bibnamefont {Yada}},\ and\ \bibinfo
  {author} {\bibfnamefont {T.}~\bibnamefont {Yokoyama}},\ }\bibfield  {title}
  {\bibinfo {title} {Topology of andreev bound states with flat dispersion},\
  }\href {https://doi.org/10.1103/PhysRevB.83.224511} {\bibfield  {journal}
  {\bibinfo  {journal} {Phys. Rev. B}\ }\textbf {\bibinfo {volume} {83}},\
  \bibinfo {pages} {224511} (\bibinfo {year} {2011})}\BibitemShut {NoStop}%
\bibitem [{\citenamefont {Sato}\ and\ \citenamefont
  {Fujimoto}(2010)}]{PhysRevLett.105.217001}%
  \BibitemOpen
  \bibfield  {author} {\bibinfo {author} {\bibfnamefont {M.}~\bibnamefont
  {Sato}}\ and\ \bibinfo {author} {\bibfnamefont {S.}~\bibnamefont
  {Fujimoto}},\ }\bibfield  {title} {\bibinfo {title} {Existence of majorana
  fermions and topological order in nodal superconductors with spin-orbit
  interactions in external magnetic fields},\ }\href
  {https://doi.org/10.1103/PhysRevLett.105.217001} {\bibfield  {journal}
  {\bibinfo  {journal} {Phys. Rev. Lett.}\ }\textbf {\bibinfo {volume} {105}},\
  \bibinfo {pages} {217001} (\bibinfo {year} {2010})}\BibitemShut {NoStop}%
\bibitem [{\citenamefont {Tanaka}\ \emph {et~al.}(2010)\citenamefont {Tanaka},
  \citenamefont {Mizuno}, \citenamefont {Yokoyama}, \citenamefont {Yada},\ and\
  \citenamefont {Sato}}]{PhysRevLett.105.097002}%
  \BibitemOpen
  \bibfield  {author} {\bibinfo {author} {\bibfnamefont {Y.}~\bibnamefont
  {Tanaka}}, \bibinfo {author} {\bibfnamefont {Y.}~\bibnamefont {Mizuno}},
  \bibinfo {author} {\bibfnamefont {T.}~\bibnamefont {Yokoyama}}, \bibinfo
  {author} {\bibfnamefont {K.}~\bibnamefont {Yada}},\ and\ \bibinfo {author}
  {\bibfnamefont {M.}~\bibnamefont {Sato}},\ }\bibfield  {title} {\bibinfo
  {title} {Anomalous andreev bound state in noncentrosymmetric
  superconductors},\ }\href {https://doi.org/10.1103/PhysRevLett.105.097002}
  {\bibfield  {journal} {\bibinfo  {journal} {Phys. Rev. Lett.}\ }\textbf
  {\bibinfo {volume} {105}},\ \bibinfo {pages} {097002} (\bibinfo {year}
  {2010})}\BibitemShut {NoStop}%
\bibitem [{\citenamefont {Morimoto}\ and\ \citenamefont
  {Furusaki}(2014)}]{PhysRevB.89.235127}%
  \BibitemOpen
  \bibfield  {author} {\bibinfo {author} {\bibfnamefont {T.}~\bibnamefont
  {Morimoto}}\ and\ \bibinfo {author} {\bibfnamefont {A.}~\bibnamefont
  {Furusaki}},\ }\bibfield  {title} {\bibinfo {title} {Weyl and dirac
  semimetals with ${Z}_{2}$ topological charge},\ }\href
  {https://doi.org/10.1103/PhysRevB.89.235127} {\bibfield  {journal} {\bibinfo
  {journal} {Phys. Rev. B}\ }\textbf {\bibinfo {volume} {89}},\ \bibinfo
  {pages} {235127} (\bibinfo {year} {2014})}\BibitemShut {NoStop}%
\bibitem [{\citenamefont {Chang}\ and\ \citenamefont
  {Coleman}(2018)}]{PhysRevB.97.155134}%
  \BibitemOpen
  \bibfield  {author} {\bibinfo {author} {\bibfnamefont {P.-Y.}\ \bibnamefont
  {Chang}}\ and\ \bibinfo {author} {\bibfnamefont {P.}~\bibnamefont
  {Coleman}},\ }\bibfield  {title} {\bibinfo {title} {Parity-violating
  hybridization in heavy weyl semimetals},\ }\href
  {https://doi.org/10.1103/PhysRevB.97.155134} {\bibfield  {journal} {\bibinfo
  {journal} {Phys. Rev. B}\ }\textbf {\bibinfo {volume} {97}},\ \bibinfo
  {pages} {155134} (\bibinfo {year} {2018})}\BibitemShut {NoStop}%
\bibitem [{\citenamefont {Wan}\ \emph {et~al.}(2011)\citenamefont {Wan},
  \citenamefont {Turner}, \citenamefont {Vishwanath},\ and\ \citenamefont
  {Savrasov}}]{PhysRevB.83.205101}%
  \BibitemOpen
  \bibfield  {author} {\bibinfo {author} {\bibfnamefont {X.}~\bibnamefont
  {Wan}}, \bibinfo {author} {\bibfnamefont {A.~M.}\ \bibnamefont {Turner}},
  \bibinfo {author} {\bibfnamefont {A.}~\bibnamefont {Vishwanath}},\ and\
  \bibinfo {author} {\bibfnamefont {S.~Y.}\ \bibnamefont {Savrasov}},\
  }\bibfield  {title} {\bibinfo {title} {Topological semimetal and fermi-arc
  surface states in the electronic structure of pyrochlore iridates},\ }\href
  {https://doi.org/10.1103/PhysRevB.83.205101} {\bibfield  {journal} {\bibinfo
  {journal} {Phys. Rev. B}\ }\textbf {\bibinfo {volume} {83}},\ \bibinfo
  {pages} {205101} (\bibinfo {year} {2011})}\BibitemShut {NoStop}%
\bibitem [{\citenamefont {Nielsen}\ and\ \citenamefont
  {Ninomiya}(1981)}]{NIELSEN198120}%
  \BibitemOpen
  \bibfield  {author} {\bibinfo {author} {\bibfnamefont {H.}~\bibnamefont
  {Nielsen}}\ and\ \bibinfo {author} {\bibfnamefont {M.}~\bibnamefont
  {Ninomiya}},\ }\bibfield  {title} {\bibinfo {title} {Absence of neutrinos on
  a lattice: (i). proof by homotopy theory},\ }\href
  {https://doi.org/https://doi.org/10.1016/0550-3213(81)90361-8} {\bibfield
  {journal} {\bibinfo  {journal} {Nuclear Physics B}\ }\textbf {\bibinfo
  {volume} {185}},\ \bibinfo {pages} {20} (\bibinfo {year} {1981})}\BibitemShut
  {NoStop}%
\bibitem [{\citenamefont {Armitage}\ \emph {et~al.}(2018)\citenamefont
  {Armitage}, \citenamefont {Mele},\ and\ \citenamefont
  {Vishwanath}}]{RevModPhys.90.015001}%
  \BibitemOpen
  \bibfield  {author} {\bibinfo {author} {\bibfnamefont {N.~P.}\ \bibnamefont
  {Armitage}}, \bibinfo {author} {\bibfnamefont {E.~J.}\ \bibnamefont {Mele}},\
  and\ \bibinfo {author} {\bibfnamefont {A.}~\bibnamefont {Vishwanath}},\
  }\bibfield  {title} {\bibinfo {title} {Weyl and dirac semimetals in
  three-dimensional solids},\ }\href
  {https://doi.org/10.1103/RevModPhys.90.015001} {\bibfield  {journal}
  {\bibinfo  {journal} {Rev. Mod. Phys.}\ }\textbf {\bibinfo {volume} {90}},\
  \bibinfo {pages} {015001} (\bibinfo {year} {2018})}\BibitemShut {NoStop}%
\bibitem [{\citenamefont {B\'eri}(2010)}]{PhysRevB.81.134515}%
  \BibitemOpen
  \bibfield  {author} {\bibinfo {author} {\bibfnamefont {B.}~\bibnamefont
  {B\'eri}},\ }\bibfield  {title} {\bibinfo {title} {Topologically stable
  gapless phases of time-reversal-invariant superconductors},\ }\href
  {https://doi.org/10.1103/PhysRevB.81.134515} {\bibfield  {journal} {\bibinfo
  {journal} {Phys. Rev. B}\ }\textbf {\bibinfo {volume} {81}},\ \bibinfo
  {pages} {134515} (\bibinfo {year} {2010})}\BibitemShut {NoStop}%
\bibitem [{\citenamefont {Fang}\ and\ \citenamefont
  {Fu}(2019)}]{doi:10.1126/sciadv.aat2374}%
  \BibitemOpen
  \bibfield  {author} {\bibinfo {author} {\bibfnamefont {C.}~\bibnamefont
  {Fang}}\ and\ \bibinfo {author} {\bibfnamefont {L.}~\bibnamefont {Fu}},\
  }\bibfield  {title} {\bibinfo {title} {New classes of topological crystalline
  insulators having surface rotation anomaly},\ }\href
  {https://doi.org/10.1126/sciadv.aat2374} {\bibfield  {journal} {\bibinfo
  {journal} {Science Advances}\ }\textbf {\bibinfo {volume} {5}},\ \bibinfo
  {pages} {eaat2374} (\bibinfo {year} {2019})}\BibitemShut {NoStop}%
\end{thebibliography}%

\end{document}